\documentstyle[12pt,aaspp4]{article}

\begin{document}

\title
    {  Standard Solar Models in the Light of New Helioseismic Constraints\\    
I The Solar Core}
    
\author{A. S. BRUN, S. TURCK-CHIEZE}

\affil{CEA/DSM/DAPNIA/Service d'Astrophysique, CE Saclay,
 91191 Gif-sur-Yvette Cedex 01, France\\}
 
\and
 
\author{P. MOREL}

\affil{D\'epartement Cassini, OCA, BP 4229, 06304 Nice Cedex 4, France\\}

\begin{abstract}
In this paper, we examine a new updated solar model taking 
advantage of the recent reexamination of the nuclear reaction rates and of
the microscopic diffusion of helium and heavy elements. Our best model 
fits  the helioseismic data reasonably well: the base of the convective zone is at
 $R_{bcz}=0.715$, the  photospheric helium 
in mass fraction is 0.243, the  sound speed square difference between the 
Sun and the model, 
 $\delta c^2/c^2<1\%$ . This model 
 leads to a reestimate of neutrino fluxes: 7.18  SNU for the chlorine 
 experiment,  127.2 SNU for the gallium detector and 4.82 $ 10^6  \rm cm^{-2} s^{-1}$
  for the $^8$B neutrino flux. Acoustic 
 mode predictions are also estimated.
 
  We then consider the radiative zone and discuss  what we learn from 
 such a model confronted with the present helioseismic constraints from 
 space experiments aboard SOHO. We present three models which respect 
 these constraints and fit the seismic observations better by taking
  advantage of the known physical uncertainties: 
 nuclear reaction rates, CNO abundances and microscopic diffusion.
We also study some current questions as the possibility of mixing in the 
nuclear core,
the revision of the solar radius, or the influence of the solar age.

   We conclude that the standard model, inside its 
 inherent uncertainties, is robust in the light of the present acoustic mode detection
 and that mixing in the core is not really favoured, even though 
 a proper understanding of the angular momentum evolution with time has not 
 yet been obtained. 
 
  The initial solar helium abundance seems more and more constrained; 
  this study supports 
 an initial abundance between 0.273 and 0.277 in mass fraction.

 This analysis allows us to define minimal values for neutrino predictions, compatible
  with present seismic results.
  We notice that a reduction of about $30\%$ on chlorine and water detectors, 
  which is more than half of the discrepancy 
 with the experimental results, 
  is still 
 supported by the present study. This work emphasizes also the fact that 
 acoustic mode
 determination does not put strong constraints on the nuclear plasma characteristics.
 
 We finally estimate g-mode frequencies in the range which may be accessible
  to the satellite SOHO,  these results emphasize the substantially improved 
  sensitivity of these modes on 
 the details 
 of the nuclear solar core, and show the frequency dependence of these modes for 
 the different models previously discussed.
 \end{abstract}
 

\section{INTRODUCTION}

Standard solar models, and stellar evolution in general, have been substantially improved in recent years, taking advantage
of the impressive helioseismic tool. The quality and accuracy of this technique
has revealed that the Sun may be considered  a real laboratory to check the physical processes
we introduce in  calculations: equation of state, radiation transport and
gravitational settling of elements.

Effectively the extraction of the internal radial sound speed from a 
great number of acoustic
 modes  determined by ground networks GONG (Hill et al. 1994, 1996), 
 IRIS (Grec et al. 1991, Gelly et al. 1997), BiSON (Elsworth et al. 1994, 
 Chaplin et al. 1996a, b) and 
 LOWL (Tomczyk et al. 1995), 
  has allowed the determination of 
 the base of the convection zone ( $r_{bcz} = 0.713 \pm 0.003 R_{\odot}$: 
 Christensen-Dalsgaard, Gough and Thompson 1991, Basu and Antia 1997), 
 the verification of the equation of state for light elements,
 with the precise determination of photospheric helium content $Y_{ph}= 0.249 \pm 0.003$
 ( Vorontsov, Baturin and Pamyatnykh 1992, Basu and Antia 1995), 
 thus compelling the 
 introduction of the slow microscopic diffusion process.
 
 Three additional space helioseismic experiments aboard SOHO (Domingo et al. 1995), 
 GOLF 
 (Gabriel et al. 1995, 1997), VIRGO (Fr\"{o}hlich et al. 1995, 1997) and MDI 
 (Scherrer et al. 1995, Kosovichev et al. 1997)  give us 
 new constraints on the transition region between radiation  and 
 convection zones and on the nuclear core, through a detailed analysis of the 
 rotation and sound speed profiles. In particular, 
 the GOLF experiment, accessing  more 
 than 100 modes penetrating the solar core, will hopefully
  constrain the nuclear energy production region down to 0.05 $R_{\odot}$ 
  and its evolution with time, if any (Lazrek et al. 1997, Turck-Chi\`eze et al. 1997). 
 
 In parallel,  solar neutrino detection has been substantially improved and 
 enlarged with GALLEX 
 (Hampel et al. 1996 and references therein) and SAGE (Abdurashitov et al. 1994,
  1996) results, sensitive to pp neutrinos. Additionally, the very recent 
 SuperKamiokande (Suzuki 1994, Totsuka et al. 1996) experiment has detected,
 for the first time, 4500 solar neutrinos 
 in less than 1 year. The next years in this field are also extremely promising 
 with the beginning of SNO (McDonald 1995) and BOREXINO (Raghavan 1995).

 This exceptional circumstance opens new debates on stellar modeling 
 aspects and on 
 neutrino predictions. All the measurements together,
  for at least half of a solar cycle, provide a unique occasion 
 to investigate models beyond  
 the standard stellar framework and to bring some quantitative answers
 to dynamical aspects which may induce turbulence and the  mixing of elements
 in different parts of the Sun or more generally local instabilities. 
 Another important issue could be a determination of specific
  properties of  neutrinos which 
 are not accessible in the laboratory.
 
In this paper, we first discuss the improved microscopic physics of 
our standard model, including the recent updated  nuclear reaction rates 
which are fundamental in the core description  and the microscopic diffusion 
 (section 2). 
We describe the reference model obtained with the stellar evolution code, 
CESAM, developed by P. Morel (1997) which is numerically sufficiently accurate 
to discuss the new generation of experimental results (better than $10^{-4}$). 
We compare our predictions on neutrino fluxes, sound speed profile and
low degree acoustic modes (using the 
pulsation code of J. Christensen Dalsgaard, 1982),
 to measurements or observations (section 3). In section 4, the differences are examined and a new step 
 consists of discussing some uncertainties of the physical processes already included
 in order to better match the helioseismic constraints. Then, we question
 some other processes as  mixing processes, radius determination or solar age. 
 Section 5
 is devoted to a general discussion and to gravity mode predictions supported 
 by the present acoustic mode
 observations. Finally section 6 summarizes our results.

\section{PHYSICAL INPUTS}

\subsection{Composition and Opacity Coefficients}

Chemical composition plays a crucial role in solar modeling, mainly via  
the opacity coefficients and the nuclear reaction rates. 
Improvements on photospheric abundances have been made since 
the review of Anders and Grevesse (1989), suppressing the disagreement
 between photospheric and meteoritic iron
and improving the knowledge of carbon, nitrogen and oxygen  
photospheric abundances (Grevesse and Noels 1993).
 The present solar photospheric metallicity agrees  very well now
 with the meteoritic composition but the present accurary does not exclude 
 a small effect of diffusion between the initial composition and 
 the present photospheric observation. Effectively, this study has led 
 to a ratio Z/X = 0.0245 but
 with an uncertainty of about 10-15 $\%$ which seems largely dominated by CNO 
 determination. 
  The first impact of these improvements on the solar model has been studied 
 by Turck-Chi\`eze and Lopes (1993).
 
  Other crucial progress
 has been achieved by the helioseismic community in the  determination
 of the photospheric 
  helium content.
   Until recently, the solar initial helium content, 
   thought to be equal to the photospheric abundance,
  was deduced from solar evolution in the absence 
  of photospheric line measurements.
 The possibility of reaching 
 a precise photospheric value from helioseismic determination 
 of the adiabatic exponent,  has been useful complementary information.
 This value of $0.25 \pm 0.01$ (Vorontsov, Baturin and Pamyatnykh 1992) or  $0.249 \pm 0.003$ 
 (Basu and Antia 1995)   is
 not very far from  
 cosmological  and certainly smaller than the initial solar value deduced 
 from solar modeling. This result has 
 largely confirmed the need to 
 introduce microscopic diffusion in stellar modeling (see section 2.4). 
 
 Despite all these improvements, there are still some elements such as $^7$Li 
 and $^9$Be, for which the observed
  surface abundance depletion is difficult to reproduce, 
 without invoking other physical processes than the ones introduced in the classical picture 
 such as mass loss by winds or turbulent mixing.\\
 
 The CESAM code used in this study, follows the time evolution from the 
 zero-age main sequence (ZAMS) or the pre-main sequence (PMS), of 12 chemical 
 elements, namely: $^1$H, $^2$H, $^3$He, $^4$He, $^7$Li, $^7$Be, $^{12}$C, 
 $^{13}$C, $^{14}$N, $^{15}$N, $^{16}$O, $^{17}$O. 
 Apart for the study of $^7$Li depletion, the impact of the PMS on the solar
 structure or the neutrinos fluxes, is negligible.\\
 The most recent observed solar abundances (Grevesse and Noels 1993)
  are used in the models described below, 
  after normalization of the metals to 1 (see Table \ref{table 1}). 
  For the $^3$He case, starting on the zero-age main sequence, 
  we use the value of $^3$He/$^1$H $=4.4 \pm 1.5 \times 10^{-5} $ 
  including deuterium burning, discussed
  in Turck-Chi\`eze et al. (1993). This value is not ruled out
  by the recent reestimate of  $^2$H/$^1$H abundance of
  Gautier and Morel (1997).\\

\begin{table}[htbp]
\caption[]{\label{table 1} Metal Abundances}
\vspace{-0.1cm}
\begin{center}
\begin{tabular}{lll}
\hline 
Element & Relative Number Fraction &  \\
\hline
\vspace{-0.2cm} \\
C & 0.24551 &\\
N & 0.06458 &\\
O & 0.51295 &\\
Others & 0.17696 &\\
\hline
\end{tabular}
\end{center}
\end{table} 

The Livermore opacity tables OPAL96 (Iglesias and Rogers 1996) 
are also based on Grevesse and Noels 
 (1993) composition and now include 19 heavy elements
 to properly introduce  the respective composition 
 of the heavy elements in these calculations. 
 The tables
 have been introduced in the CESAM code and  the mean Rosseland opacity, $\kappa_R$ 
 is calculated for each mass shell, interpolating across the tables for  X, T, 
 $\rho$ and Z. 
 One of the difficulties in using such tables is to extend them 
 outside their original range. We accomplished this by using Kurucz's low temperature 
 opacities below T$_{inf}=5700K$. Specific studies have been performed
 with higher matching temperatures (up to T$=10^4K$) to check the atmospheric 
 T($\tau$) laws (Brun et al. 1997). The smoothest 
 transition was obtained for a temperature of T$=9300K$ 
 (the discrepancy between these two calculations is smaller than 2\%), even though
 the very small molecular components encourage the use of the Livermore tables down to the
 photospheric temperature (Sharp and Turck-Chi\`eze, 1998).\\
 The recent updated OPAL opacities may introduce up to a 10\% change 
 in the opacity coefficients due to the improvement of both the physical 
 description, the numerical procedures and the inclusion of 19 instead of 12 
 heavy elements. A comparison between OPAL93 (Iglesias et al. 1992) and OPAL96 
 opacity tables indicates variations up to 5\%  
 in the intermediate region of the solar models 
 (0.3 R$_{\odot}$ - 0.7 R$_{\odot}$). Such modifications have a clear
 effect on the sound speed profile at the level of accuracy we reach today 
 (i.e few \% in opacity $\rightarrow$ few $10^{-3}$ in sound speed) and 
 leads to a slight degradation when comparing with the helioseismic data. 
 These opacity modifications explain partly why solar models using 
 older opacities, like model S of Christensen-Dalsgaard et al. (1996), have
  slightly different sound speed profiles. We have verified that our result 
  is independent
  of the opacity interpolation subroutine (the subroutine delivered by the 
  Livermore
  group leads to a radial sound speed profile in agreement within 0.001 
  with the one obtained in using the opacity  interpolation subroutine provided 
  by G. Houdek 1996). The other sources of difference 
  are attributed to the used nuclear reaction rates (see 2.3) and
  the solar age used for the computation (see 4.4.2). \\ 

\subsection{Equation of State}

We have introduced the Livermore EoS in the CESAM code. This equation of state
considers the
correct ionization treatment of the chemical species present in the solar 
plasma, pressure 
ionization and quantum corrections other than electronic degeneracy 
(Rogers, Swenson and Iglesias 1996).  Moreover it is consistent
 with the opacity coefficients. 

\begin{figure}[htbp]
\vspace{2.cm}
\setlength{\unitlength}{1.0cm}
\begin{picture}(9.5,6.5)
\includegraphics{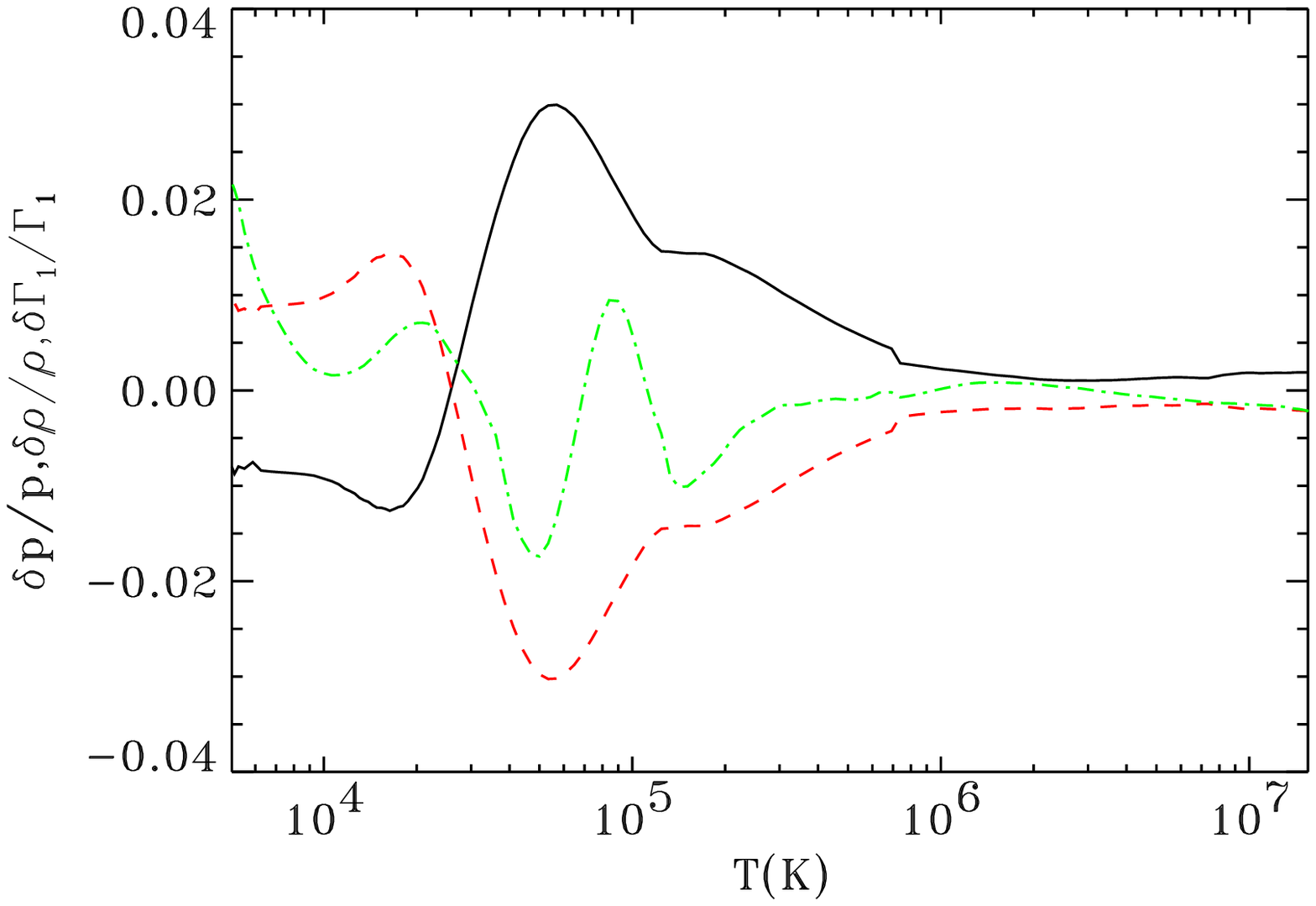}
\end{picture}
\vspace{-0.2cm}
\noindent
\begin{center}
Fig 1: Comparison of the main thermodynamical quantities: 
 the pressure P, the density $\rho$, and the adiabatic exponent
 $\Gamma_1$ for the two equations of state: VDC and OPAL96  in the sense
  (VDC-OPAL)/OPAL. $\delta P/ P$ -----, $\delta \rho / \rho$ - - - 
  and $\delta \Gamma_1 / \Gamma_1$ -.-.- .
  \end{center}
\vspace{0.5cm}
\end{figure}  
 
 In figure 1 we compare this equation of state with the one used 
 in our previous solar models (Turck-Chi\`eze et al. 1988, 
 Turck-Chi\`eze and Lopes 1993, Dzitko et al. 1995).
  It was
 based on  a treatment of a perfect gas for ions and  an estimate of the partial 
 ionization of light elements using the Saha equation (Vardya 1960), 
 a correct estimate of the electron degeneracy
  and a coulomb term recently reestimated beyond 
 the Debye approximation (Dzitko et al. 1995). We call it  V (Vardya) 
 D (degeneracy)
 C (coulomb correction) equation of state in the following. 
 This equation had 
 the advantage of being analytical, which allows an estimate of the different terms
 and an easy derivation of each of them,  but the 
 partial ionization of the heavy elements was not considered and it
 was not  accurate enough to 
 estimate the partial ionization of the light elements, even if 
 molecular components were included. The oscillatory behaviour of 
 $\delta \Gamma_1 \over \Gamma_1$ is due to the different locations 
 of the partial ionization zones of hydrogen and helium.
 In fact the improvements of the Livermore equation of state are
 effectively mainly in this zone corresponding to
  T$<5 \times 10^5$K.
 The small difference  of $0.5\%$ in the central region
 could be due to the complete ionization of all the species assumed in VDC 
 EoS but may be  also due to a slight difference in the
 coulomb 
 correction for intermediate plasma coupling. 

Due to the small role of the heavy elements in the equation of state,
 we have only calculated models with OPAL 
EoS at fixed Z element composition. Most of the models used Z$=0.0195$. 
Guzik et al. (1995) show that the effect of a fixed Z  
in the EoS was observable ($< 1 \mu$Hz) in the acoustic mode frequencies. 
On the contrary, the opacity $\kappa_R$ has been evaluated at each point 
with a precise Z composition (see section 2.4), as the heavy elements, and more specifically
$^{16}$O and $^{56}$Fe 
are strong contributors to the mean Rosseland opacities (Courtaud et al. 1990).

Due to its lower temperature boundary T$_{inf}=5000K$, 
(the minimal temperature value in our models is $\sim$ 4200K),
 Livermore EoS could not be used along the whole structure.
   So VDC EoS was used  for the very upper layers. 
   This absence of consistency creates a small
   discontinuity in pressure, 
   and a difference of about $1 \mu$Hz on acoustic-mode frequencies at high frequency. 
\\
\subsection{Nuclear reaction rates}

The nuclear reaction rates are the most important ingredients in the 
description of 
the solar nuclear core and for the prediction of the neutrino fluxes, mainly for secundary 
neutrinos as those of the ppII, ppIII chains and CNO cycle 
(Turck-Chi\`eze and Brun 1997). This is the reason 
why considerable effort is still devoted to the best determination 
of the different reaction interactions, in improving the theoretical 
determination of the electronic capture on $^7$Be (Gruzinov and Bahcall 1997), 
the experimental conditions for ($^3$He,$^3$He)
(Arpesella et al. 1996) and ($^7$Be,p) (Hammache et al. 1998),
 the extrapolation of measurements to the solar range of energy,
 the effect of 
solar plasma screening on the reaction rates (Turck-Chi\`eze and Lopes 1993, 
Dzitko et al. 1995). 

The importance of these ingredients in  the solar model calculation
has led some of us
to properly examine all the progress obtained and determine a new 
update
of all the reaction rates and screening effects, including recent works 
and commenting the ones which have been
excluded in order to get updated nuclear S factors and corresponding 
uncertainties 
(Adelberger et al. 1998). 

This updated compilation, including Mitler screening prescription
(Dzitko et al. 1995), has been
introduced in the CESAM code in place
  of Caughlan and Fowler (1988) rates and weak screening (Salpeter 1954). 
  The new values for the cross section factor S(0) and its derivatives 
  are listed in Table \ref{table 2}, for the main nuclear
   reaction rates of the pp chain. 
 
\begin{table}[htbp]
\caption[]{\label{table 2} Nuclear S factors and derivatives 
for the main reactions of the hydrogen burning (see Aldelberger et al. 
1998 for the others) and screening factors for weak or intermediate screening.}
\vspace{-0.1cm}
\begin{center}
\begin{tabular}{llllll}
\hline 
Reaction & S(0) & S'(0) & S''(0) & $\rm f_{WS}$ & $\rm f_{Mitler}$ \\
         & (MeV barn)& (barn) & (b MeV$^{-1}$) & \\
\hline
\vspace{-0.2cm} \\
$^1$H(p,e$^+$ $\nu$)$^2$H & 4.00 10$^{-25} $& 4.48 10$^{-24}$& & 1.05  & 1.045 \\
$^3$He($^3$He,2p)$^4$He &  5.4& -4.1 & 4.6&  1.21 & 1.18\\
$^3$He($\alpha$,$\gamma$)$^7$Be & 5.3 10$^{-4}$& -3.0 10$^{-4}$ & & 1.21 & 1.18\\
$^7$Be(p,$\gamma$)$^8$B & 1.9 10$^{-5}$& -1.35 10$^{-5}$& 7.33 10$^{-5}$& 1.21 & 1.17\\
$^{14}$N(p,$\gamma$)$^{15}$O & 3.5 10$^{-3}$ & -0.0128 & &1.40 & 1.29\\
\hline
\end{tabular}
\end{center}
\end{table}

If  this table is compared to the previous recommendations, three 
main changes are apparent. The recommended  $S_{pp}(0)$ value 
is between that suggested by 
Kamionkowski and Bahcall (1994)
and that used by Turck-Chi\`eze and Lopes (1993). It includes a 
reestimate of the neutron lifetime of $888 \pm 
3$ s (Barnett et al. 1996). The present uncertainty 
 of 2.2 $\%$, is dominated by the meson current effect. The 
($^3$He,$^3$He) reaction rate has been slightly increased due 
to the recent remeasurement 
of this cross section in the laboratory
down to 12 keV without revealing any resonance. 
  The error bar of $\pm 8\%$ is mainly due to the systematic 
error
and the uncertainty on the effect of screening in the laboratory. The situation of 
$(^3He,^4He)$ is still confusing as the adopted reaction rate is the 
mean value of those of two types of experiments leading to two different results.
So this reaction rate may be reduced in the future
as the direct measurements lead to a smaller value than the $^7Be$ 
activity measurements. Finally, the 
reestimate of the ($^7$Be,p) reaction rate agrees with the analysis of Turck-Chi\`eze et al.
 (1993)
and reduces the S(0) recommended by Bahcall and Pinsonneault 
(1995) by 15$\%$.  The difficulty of this experiment due to the 
radioactive target leads to an error bar of at least $10\%$. 
The measurement of Filippone et al. (1983)
 is now confirmed by the very recent experiment of
Hammache et al. (1998). This experiment will be continued at low energy. Effectively,
a second problem for this reaction rate is the 
extrapolation at the solar 
energy of 20 keV. The calculation is still extremely uncertain, depending on 
the  way $^7$Be and $^8$B nuclei are described. A ``prudent conservative range''
 for this astrophysical factor of $19^{+8}_{-4} eV b$ is consequently recommended.

Several papers have been published on the stellar screening effect. 
Discussions during the preparation of the compilation have led to a
recommended extension of the weak screening to treat intermediate plasma
for reaction between two reactants of Z greater than 1. Following our previous study
(Turck-Chi\`eze and Lopes 1993; Dzitko et al. 1995), we  use in the present paper
 the formalism 
for weak screening for the pp interaction and intermediate screening 
based on Mitler's work (Mitler 1977) .

We recall that this
 intermediate screening (see table 2) reduces 
 the neutrino fluxes in comparison with weak screening (except for $^1$H(p,e$^+$ $\nu$)$^2$H)
 but leads to greater values than those of Graboske et al. (1973). This last formalism 
is not adequately accurate for solar neutrino predictions and does not treat the electron 
degeneracy. Recent work (Gruzinov et Bahcall 1998) 
confirms this tendency and could even support a value nearly compatible with the weak screening
for heavy elements of Z $\ge$ 8. The  difference between these 
two recent works on screening
is smaller than the uncertainty 
of the corresponding nuclear reaction rates.

\subsection{Microscopic Diffusion}

 This slow process, which requires an atom needing more than 10$^{10}$ years 
 to move along 
 the solar radius, has been neglected in solar modeling in the past. But 
 now,
  due to the evidence  of surface element depletion by helioseismic 
  analysis of photospheric helium, this process must be introduced 
  in the standard treatment of the solar model.\\
  
Consequently, the time evolution of the mass fraction abundances $X_i$ 
of each chemical species 
 must be expressed as follows:
\begin{equation}
\frac{dX_i}{dt}=\frac{dX_i}{dt}_{nucl} +\frac{dX_i}{dt}_{diff}
=\frac{dX_i}{dt}_{nucl}
-\frac{1}{\rho r^2}\frac{d}{dr}(\rho r^2 V_iX_i) 
\end{equation}
In this expression a mass loss term and a turbulent term are still missing.
For the diffusive velocity $V_i$, the CESAM code uses the 
approximate formulae proposed by Michaud and Proffit (1993, eq. 17, 18, 19) for a $^1$H-$^4$He mixture and several trace elements.

\noindent
 Formula (19) of Michaud and Proffit (1993), describing the thermal part of the diffusive velocity 
 of a trace element i in a $^1$H-$^4$He background, is in agreement 
 with a more detailed calculation based on Burgers' equations to  within 20\%
 (Burgers 1969).

 In the CESAM code,  these formulae are applied to diffuse  the following elements, 
 $^1$H, $^2$H, $^3$He, $^4$He, $^7$Li, $^7$Be, $^{12}$C, $^{13}$C, $^{14}$N, $^{15}$N, 
 $^{16}$O, $^{17}$O and of course an extra element ''Ex'' in order to diffuse the remaining chemical composition (i.e $X_{Ex}=1-\sum_{i=H}^{O17}X_i$), treated as silicium (Morel et al. 1997).

Figure 2 shows the diffusive velocities for 
  the different elements in our reference model along with the  
  contribution of the thermal velocity (up to 40\% of the total velocity).
   It is interesting to note, that the bumps close to T$=10^7$K on $^{12}$C and $^{14}$N 
   microscopic velocities, arise from the composition gradient due to the nuclear processes 
   occuring in this region.

\begin{figure}[htbp]
\vspace{2.cm}
\setlength{\unitlength}{1.0cm}
\begin{picture}(9,15)
\includegraphics{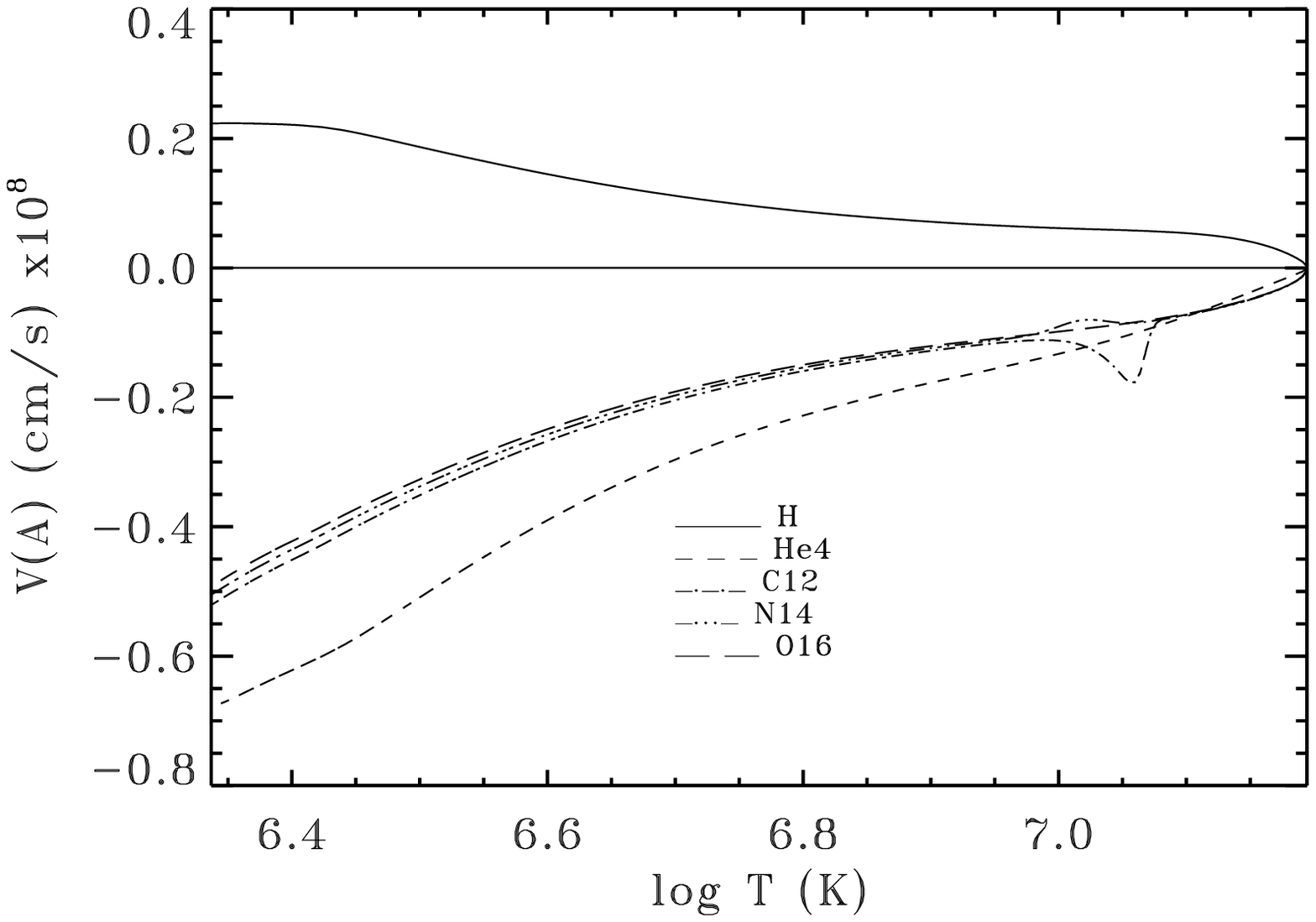}
\includegraphics{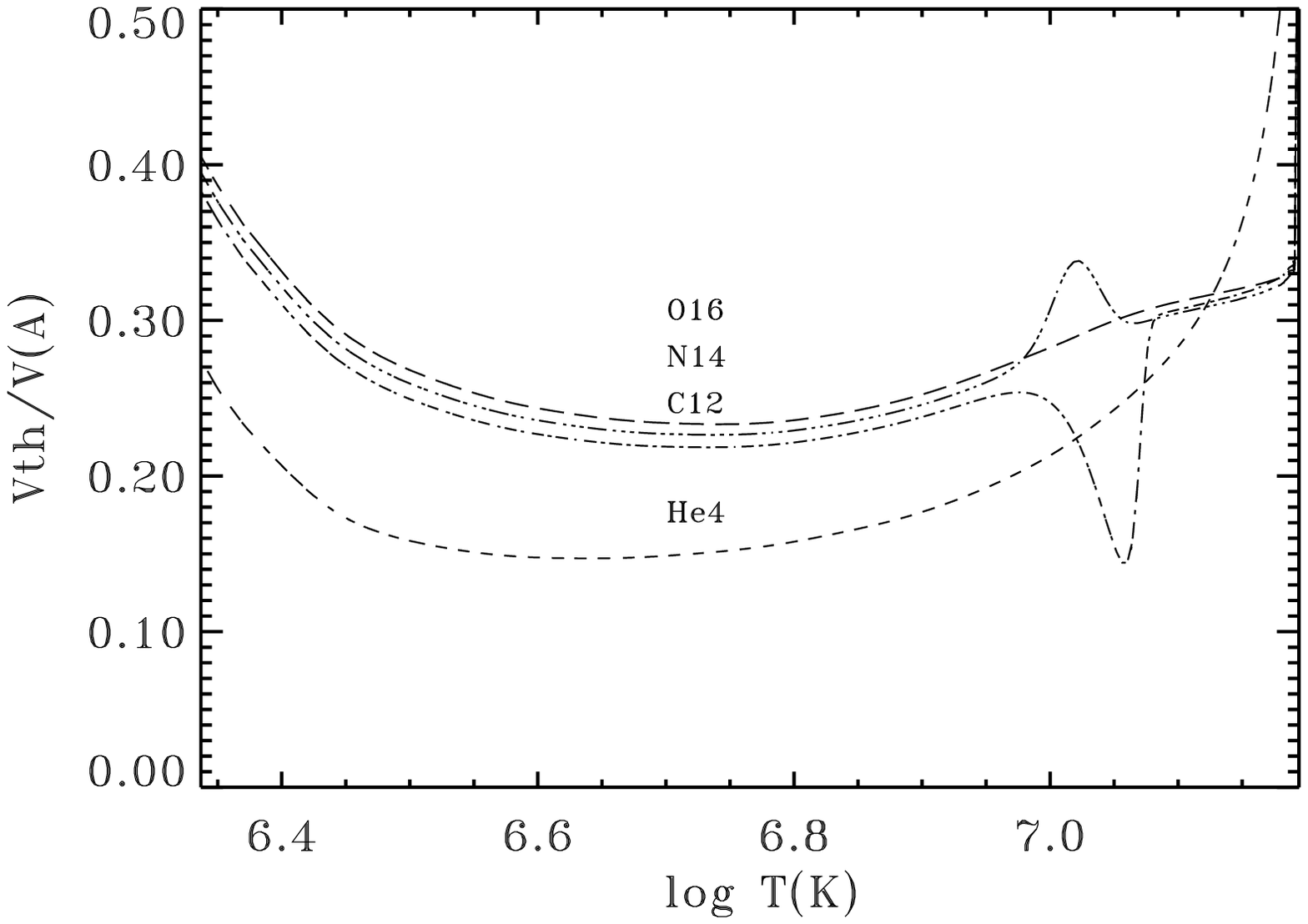}
\end{picture}
\vspace{0.2cm}
\begin{center}
Fig 2:	
Diffusive velocities and thermal contribution to these velocities for the different 
element considered.
\end{center}			
\vspace{-0.3cm}
\end{figure} 
\vspace{.5cm}
 
\begin{figure}[htbp]
\vspace{2.cm}
\setlength{\unitlength}{1.0cm}
\begin{picture}(7,15)
\includegraphics{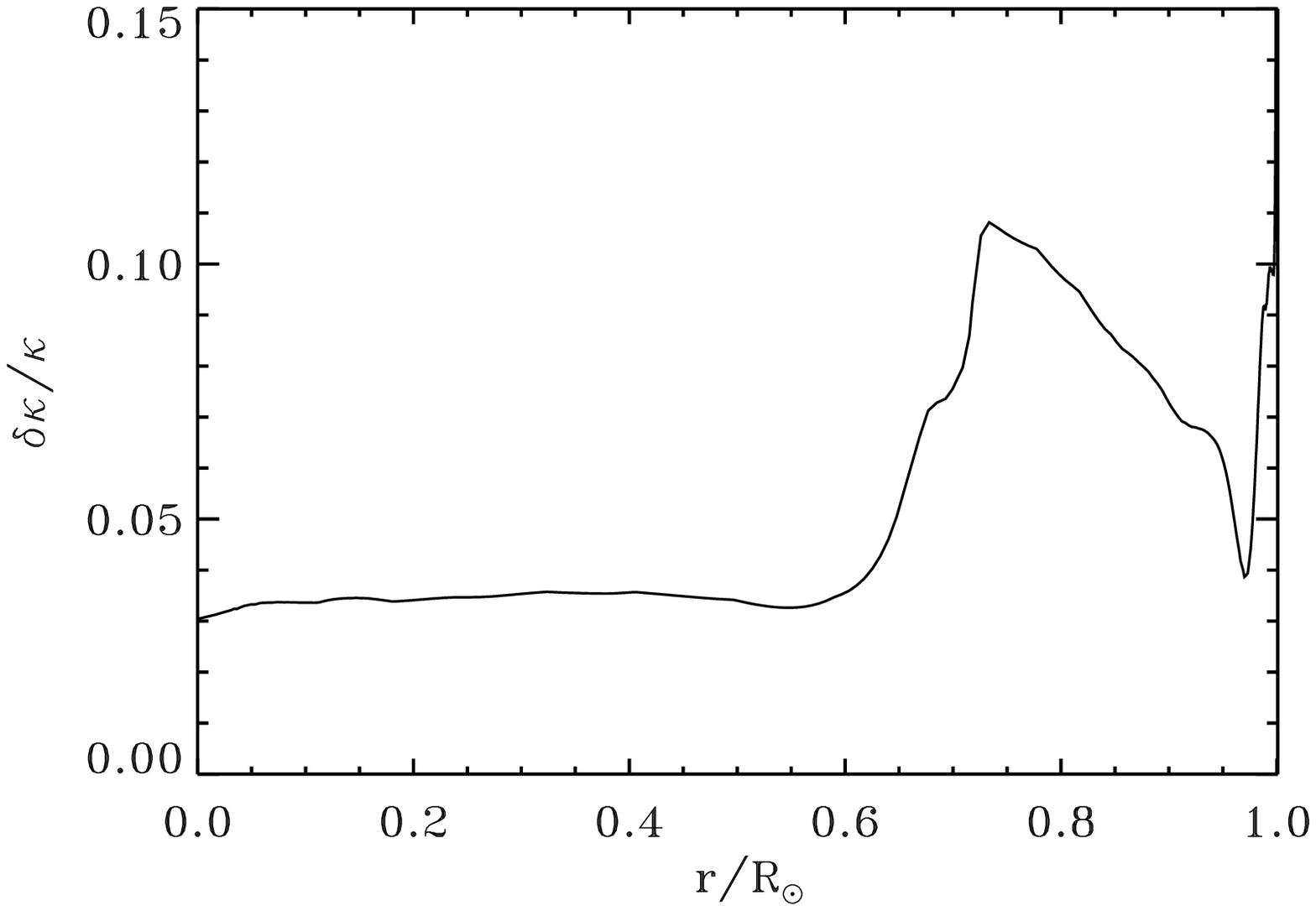}
\includegraphics{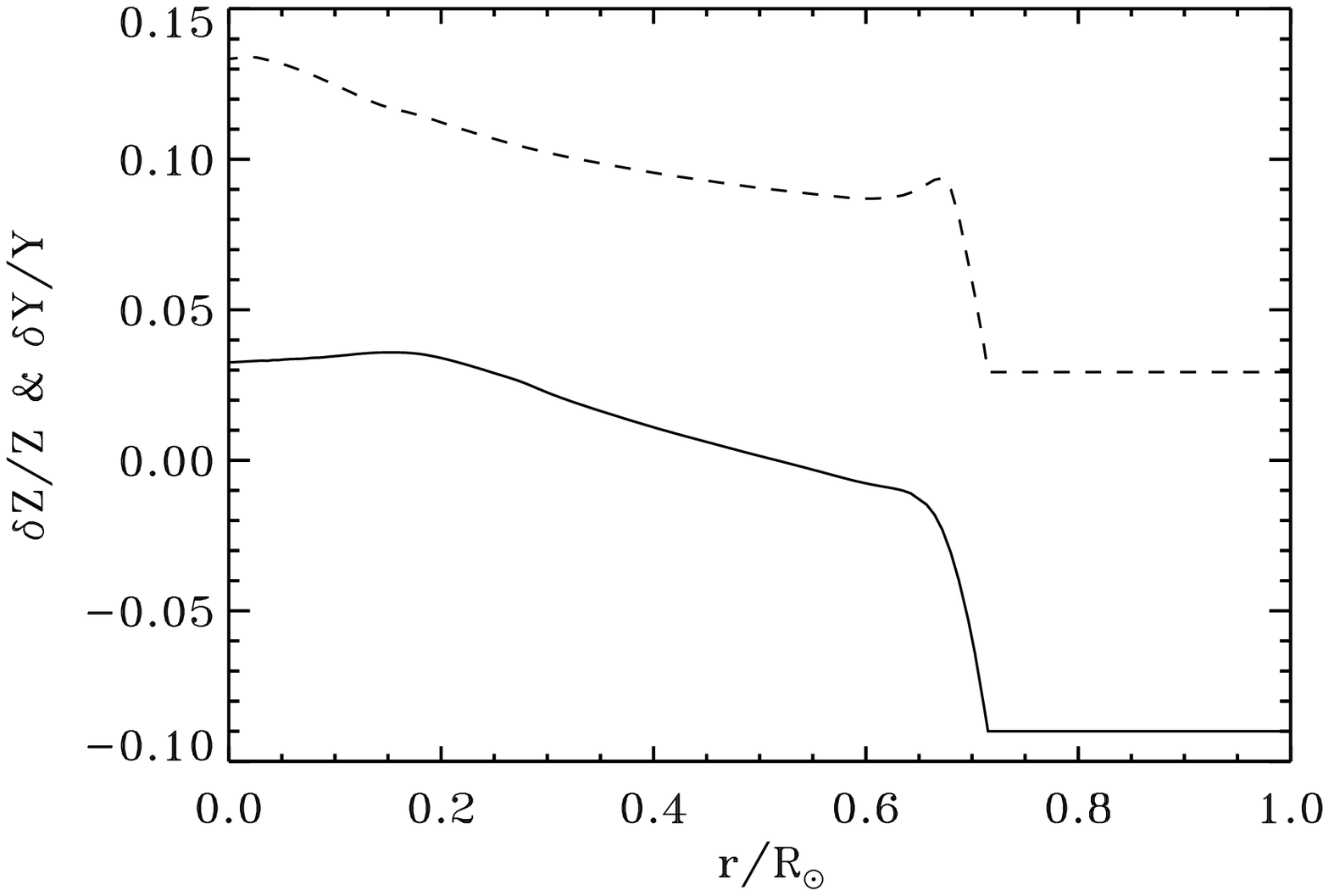}
\end{picture}
\vspace{-0.2cm}
\begin{center}
 Fig 3: a) Opacity difference between diffusive and non-diffusive models.
 b) Corresponding relative variation of helium and heavy elements.
 \end{center}			
\vspace{0.3cm}
\end{figure}

Figures 3 a) and b) clearly show the impact  
of the variation of the composition along the solar radius on the opacity coefficient. The large bump 
around the base of the convective zone shows the crucial role of the 
heavy elements in this range, mainly iron and oxygen.
As is well known, this has a direct consequence on the sound speed profile .
Due to the approximation done in this calculation, we have compared our 
results with other more recent detailed calculations of microscopic diffusion 
processes.
 We notice that the diffusion prescription we use in the present work, matches 
  a more complete description based on Burgers full equations  within 5-10\%
 (Thoul, Bahcall and Loeb 1993, Turcotte et al. 1998), and has the advantage 
 of a dramatic saving in CPU time.
 
 Turcotte et al. (1998)  have performed a very complete microscopic calculation taking 
into account the radiative acceleration using monochromatic opacity tables and 
the average atomic number to treat the fact that the chemical species are not 
completely ionized.
For example,  the photospheric helium variation, between a model without or with diffusion,
 is about 11.1\% for Turcotte et al. (1998),
 when  partial ionization is included (model C), and 10.4\% for the full 
 calculation (model H). In our model, where complete ionization has 
 been assumed for all elements, we find a 10.6\% helium variation. 
The radiative acceleration reduces the settling mainly for the 
elements of the iron peak but does not act strongly on He, C, N, O 
(less than 2\% of the gravitational acceleration except at the base 
of the convective zone where an effect 
of 4-6\% could be reached) (Turcotte et al. 1998).
 
 For heavy elements the difference is slightly greater mainly due to the expression 
 of $v_{therm}$ which underestimates this effect by about 20\% 
 (Michaud and Proffitt 1993). Effectively the Z variation is about 10.3\% 
 for Turcotte's model C but 8.5\% for model H 
 (one notes also that grad/g may reaches 0.4 for the elements of iron peak),
  and only 8.1\% for our diffusive 
 model. 
 Another part of the discrepancy should come from the neglect of
 the partial ionization of heavy elements such as O, Fe, near the convective 
 zone, while an additional small contribution to the difference may arise from the different ages used in the calculations (4.52 Gyr compared with 4.57 Gyr).

In conclusion, the equations we use lead to a in reasonably good 
agreement  with more precise and extremely time consuming
calculations in the case of the Sun. For greater mass or more evolved stars, 
this treatment may appear too simple. Nevertheless, this process is probably partly 
inhibited by turbulence at the basis of 
the convection zone. The discrepancy between the solar sound speed and 
that deduced from a model including turbulence, will give part of the answer (see our next paper: 
``II The radiation-convection transition").

As far as the neutrino predictions are concerned, there is an  underestimation 
of about 5\% compared to models of Turcotte et al. (1998) or Bahcall and Pinsonneault (1995) 
(see Turck-Chi\`eze 1996, Turck-Chi\`eze and Brun 1997 for a comparison).
 
\subsection{Convection}
The standard mixing length  theory (MLT) (B\"ohm-Vitense 1958) 
is used to describe the convective transport of the energy, taking into account
the optical width of the eddies (Henyey et al. 1965, see also Berthomieu et 
 al. 1993).
This concept is insufficient to compare with present helioseismic data 
(mainly for the absolute acoustic modes frequencies) and needs to be improved. 

\subsection{Atmospheric Treatment}

The treatment of the outer layers is important to determine the absolute 
values of the acoustic modes.
One major difference between our previous models (Turck-Chi\`eze et al. 1988, 
Turck-Chi\`eze and Lopes 1993 and Dzitko et al. 1995), based on the code of Paczynski 
 (1969) and the present ones based on the CESAM code is the atmosphere treatment.
 The Paczynski code uses the Henyey method to solve the structure equations. 
 The solution of these equations is done from the central part to some region 
 located in the convective zone and 
 from R$=$1.003 R$_{\odot}$ down to this same point. In this scheme, one then 
 minimizes the 
 discrepancies at the adjusting point and iterates the calculation until it matches
  the structure. 
In contrast, CESAM code uses a reconstructed atmosphere based on a 
T($\tau$,T$_{eff}$) law (Morel et al. 1994), deduced from the ATLAS9 
atmosphere code (Kurucz 1991), and adds it to the main structure. 
The matching point could be at different optical depth $\tau_b$. Normally, the 
atmosphere is better described, but at the transition a temperature gradient 
exists. It leads to a discontinuity
which deteriorates the quality of the mode frequency predictions (see Brun et al. 1997 and 
Turck-Chi\`eze et al. 1997). Pending improvements in the CESAM code
will address this problem.

\section{STANDARD MODEL RESULTS}

In this section, we present the results of our standard solar model, calculated 
with the CESAM code (Morel 1997), including 
all the physical quantities described above. 
In this calculation, we also use the physical parameters 
included in  table \ref{table 3}. 

\begin{table}[htbp]
\caption[]{\label{table 3} Solar Observations: physical parameters, helioseismic 
observations, solar neutrino detections}
\vspace{-0.1cm}
\begin{center}
\begin{tabular}{ll}
\hline 
\vspace{-0.2cm} \\
Physical parameters \\

M$_{\odot}$ $=$ (1.9891 $\pm$ 0.0004) $\times$ 10$^{33}$ g \\
R$_{\odot}$ $=$ (6.9599 $\pm$ 0.0002) $\times$ 10$^{10}$ cm \\
L$_{\odot}$ $=$ (3.846 $\pm$ 0.004) $\times$ 10$^{33}$ ergs.s$^{-1}$ \\
 Age = 4.52 $\pm$ 0.04 Gyr\\
(Z/X)$_{\odot}$ $=$ 0.0245 $\times$ (1 $\pm$ 0.1)\\
\\
Helioseismic observations \\

Y$_{surf}$ $=$ 0.249 $\pm$ 0.003\\
R$_{bcz}$/R$_{\odot}$ $=$ 0.713 $\pm$ 0.003 \\

\\
Solar neutrino detections \\

$^{71}$Ga $=$ 76 $\pm$ 8 SNU (cal= 0.91 $\pm 0.08$ ) for GALLEX\\
$^{71}$Ga $=$ 70 $\pm$ 8 SNU (cal= 0.95 $\pm 0.12$ ) for SAGE\\
$^{37}$Cl $=$ 2.55 $\pm$ 0.25 SNU for Homestake \\
$^8$B $=$ 2.7 $\pm$ 0.1 $\times 10^6$ cm$^2$ s$^{-1}$ for Kamiokande\\
 $^8$B $=$ 2.44 $\pm$ 0.26 $\times 10^6$ cm$^2$ s$^{-1}$ for Super Kamiokande\\
\vspace{-0.2cm} \\
\hline
\end{tabular}
\end{center}
\end{table}

The mass, radius and luminosity are
the usual  values. The present age is deduced 
from a study of the age of the earth and of  the solar system formation 
(Guenther 1989, 1992). This value supposes that we begin our calculation on 
the main sequence and is not in contradiction with the compilation of Wasserburg in
 Bahcall, Pinsonneault and Wasserburg (1995).
The heavy  element composition is deduced from the compilation of Grevesse 
and Noels (1993) and we have chosen a $^3$He mass composition of $\sim 9. 10^{-5}$
 which takes into account
the best determination of the isotopic ratio $^3$He/$^4$He and the deuterium burning 
during the premain sequence. 

We iterate our calculation to reach the solar luminosity, 
 radius and  Z/X ratio estimated by Grevesse (1996) within  $10^{-5}$ 
 by varying the solar initial helium, the ratio Z/X of the initial composition
  and the mixing length
 parameter $\alpha$.
 In doing so, we assume the relative composition of the heavy elements remains constant with time,
 which is not totally true, as  seen in the previous section where we discuss 
  microscopic diffusion in detail.

We generally consider about 400-600 mesh points (this number may be increased if needed)
 and  40 time steps with an adaptable step varying from 10 Myr to 200 Myr. 
 The numerical integration of the structure equations is based on 
the splines collocation method (de Boor 1978, Morel 1997).

We summary also in table 3,  the  results from helioseismology:
 the determination of the photospheric helium ($Y_{surf}$)
 (Vorontsov, Baturin and Pamyatnykh 1992, Basu and Antia 1995) 
and the position of the base of the convection zone 
($R_{bcz}/R_{\odot}$)(Christensen-Dalsgaard et al. 1991).
The mean values (on the period of observations) of  the three 
types of neutrino detections are also listed:
 gallium from
GALLEX (Hampel et al. 1996) and SAGE (Abdurashitov et al. 1996)  
(the calibration value is not introduced in the experimental mean value), chlorine from Homestake (Davis 1994)
and water for Kamiokande (Fukuda et al. 1996) and SuperKamiokande (Totsuka 1996).
They are directly compared with the results of our present solar model in  table 4.

\begin{table}
\caption[]{ \label{table 4} 
 \small{Thermodynamical Quantities and Neutrino Predictions of the Standard Solar Model:
$\rm \Delta R / R_{\odot}$ and $\rm \Delta L/L_{\odot}$: accuracy on R$\odot$ and L$\odot$, 
$\alpha$: mixing length parameter, Y$_0$, Z$_0$, (Z/X)$_0$:
initial helium, initial heavy element and initial ratio heavy element on hydrogen, 
Y$_s$, Z$_S$, $(Z/X)_S$: idem for photospheric compositions,  $\tau_b$ 
is the optical depth of the bottom of the atmosphere, R$_{bcz}$, 
T$_{bcz}$ are the radius and temperature at the base of the convective zone,
Y$_c$, Z$_c$, T$_c$, $\rho_c$: central helium, heavy element contents, 
central temperature and density; P$_0$: gravity mode period; 
$^{37}$Cl, $^{71}$Ga, $^8$B respective neutrino predictions for the chlorine, 
gallium and water detectors. OPAL/K 5800K means that we use OPAL96 opacities 
above  5800K and Kurucz ones below. }}     
\begin{center}
   \begin{tabular}{p{3cm}*{3}{c}}  	
    \hline
     Parameters	 &  model  without diffusion &  with diffusion\\
    \hline
     Opacities   & OPAL/K 5800K& OPAL/K 5800K\\
     Diffusion    & no &  yes   \\ 
     $\rm \Delta R/R_{\odot}$ & 2.26 $10^{-5}$& 1.57 $10^{-5}$\\
     $\rm \Delta L/L_{\odot}$ & 1.79 $10^{-5}$& 3.67 $10^{-5}$\\
     $\alpha$  & 1.713 & 1.840\\
     
     Y$_0$  & 0.265 &  0.273\\
     Z$_0$  & 1.757 $10^{-2}$&  1.964 $10^{-2}$\\
     (Z/X)$_0$ & 0.0245 & 0.0277\\
    
     Y$_s$  & 0.265 &  0.243\\
     Z$_s$  & 1.757 $10^{-2}$ & 1.810 $10^{-2}$\\
     (Z/X)$_s$ & 0.0245 & 0.0245\\
     $\tau_b$  &2 &  2\\
     
     R$_{bcz}$/R$_{\odot}$ & 0.729 &  0.715\\
     T$_{bcz} \times 10^6$ (K) & 2.055 & 2.172\\
     
     Y$_c$   & 0.614 &  0.635\\
     Z$_c$   & 1.807 $10^{-2}$& 2.084 $10^{-2}$\\
     T$_c \times 10^6$ (K) & 15.44 &  15.67\\
     $\rho_c$ (g/cm$^3$) & 147.80 & 151.85\\
     
     P$_{0}$ (mn)& 36.39 &   35.75 \\
     
     $\rm ^{37}Cl $ (SNU) & 5.65 & 7.18\\
    $\rm ^{71}Ga $  (SNU)& 119.4 &  127.2 \\
     $^8$B (10$^6$/cm$^2$/s)  & 3.66 &   4.82\\
    \hline
   \end{tabular}
  \vspace{-0.3cm}
  \end{center}
\end{table}

\begin{figure}[htbp]
\vspace{2.cm}
\setlength{\unitlength}{1.0cm}
\begin{picture}(9,12)
\includegraphics{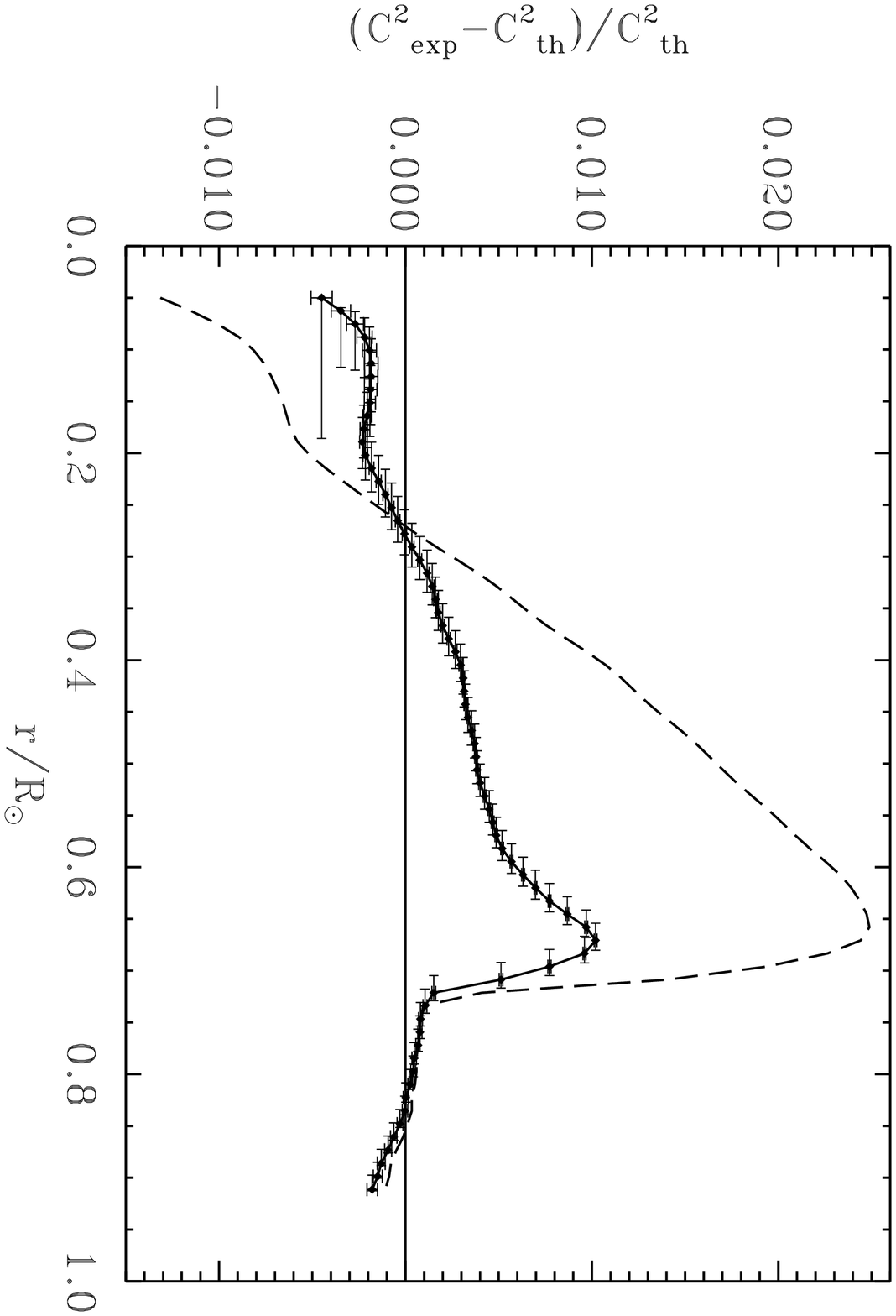}
\end{picture}
\vspace{-0.2cm}
\begin{center}
Fig 4:  Sound speed square difference between the Sun 
 measured by GOLF+LOWL experiments (Lazrek et al. 1997, Tomczyk et al.1995)
 and our models without (---) and with  diffusion (full line with 
 experimental error bars). 
 \end{center}			
\vspace{0.5cm}
\end{figure}

\begin{table*}[htbp]
\begin{center}
\caption[]{\label{table 5} Global acoustic mode frequencies ($\mu Hz$)
 obtained with our standard model for l=0,1,2,3.}
\end{center}
\vspace{-0.1cm}
\begin{tabular*}{\textwidth}{@{}l@{\extracolsep{\fill}}rrrrr}\\
\hline
   n  &  $l=0$     &      $l=1$ &       $l=2$ &      $l=3$    \\
   0  &    0.000   &   0.000  &  352.584  &  392.639\\
   1  &  258.179  &   284.435  &  382.864  &  415.854\\
   2  &  404.229   & 448.288  & 514.278   & 564.492\\
   3  &   535.571   &  596.675 &   664.025 &   718.155\\
   4  &  679.996   &  746.366  &   811.419  &   866.505\\
   5  &  824.820  &   893.258  &   959.419  &  1014.559\\
   6  &   972.161  &  1039.122  &  1104.663 &   1161.104\\
   7  &  1117.509  &  1185.051  &  1250.198  &  1306.229\\
   8  &  1262.902  &  1329.136  &  1394.069  &  1450.387\\
   9  &  1406.906  &  1472.349  &  1535.395   & 1590.863\\
  10  &  1547.873   & 1612.036   & 1673.985   & 1728.560\\
  11  & 1686.125   & 1748.765   & 1809.612  &  1864.521\\
  12  &  1821.537  &  1884.412  &  1945.228 &   2000.485\\
  13  & 1956.901   & 2020.064  &  2081.446  &  2137.350 \\
  14  &  2092.957  &  2156.401 &   2217.365 &   2273.280\\
  15  &  2228.494  &  2291.917  &  2352.670  &  2408.334\\
  16  &  2363.384  &  2426.353  &  2487.029   & 2543.238 \\
  17  &  2497.456  &  2560.885  &  2621.774 &2678.527\\
 \hline
\end{tabular*}
\end{table*} 

The  diffusion of helium and heavy 
elements
is clearly visible in the sound speed difference between the Sun and a model 
including this physical process, even though it increases the discrepancy on  neutrino fluxes 
($\sim$ +27\% for $\rm ^{37}Cl$ detector). 
This point has  already been mentioned by several groups (Proffitt and Michaud (1991), 
Bahcall and Pinsonneault (1992), 
Christensen-Dalsgaard et al. (1993), Berthomieu et al. (1993) 
for helium diffusion;
Michaud and Proffit (1994), Bahcall and Pinsonneault (1995), Morel, Provost and Berthomieu (1997) 
 for diffusion of helium + heavy elements. 
Figure 4 shows a comparison of the square of the sound speed deduced by S. Basu 
and J. Christensen Dalsgaard (Turck-Chi\`eze et al. 1997), from the 
GOLF (Lazrek et al. 1997, ) and LOWL (Tomczyk et al.
 1995) experiments compared
 with our standard models with 
and without microscopic diffusion.

  Including the effects of diffusion, 
 we observe (figure 4, table 4 and table 5):

\noindent 
a) a reduction of the discrepancy between the observed sound speed 
deduced from GOLF + LOWL data, 
and the calculated one by a factor greater than 2 when microscopic diffusion is incorporated.\\ 
b) a 10\% increase of the heavy element composition, at the center leading to
a  Y$_{surf} \sim 0.243$, in reasonable agreement with the value inferred by 
helioseismology (Basu et Antia 1995).\\
c) a much better agreement with the observed base of the convection zone for models with 
elemental diffusion : $R_{bcz}=0.715 R_\odot$ vs $R_{bcz}=0.729 R_\odot$ for 
non-diffusive models, which is due to the mean Rosseland opacity 
increase over the whole structure influenced by  C, N, O and Fe composition modification, 
with a dominant peak close to the base of the convection zone. 
This has the effect to increase $\nabla_{rad}$ and thus to favour convective energy 
transport and a broader convective zone.\\
d) a very reasonable agreement between our acoustic mode predictions and the most recent 
observations
from GOLF (Lazrek et al. 1997, Rhodes et al. 1997) for the low frequency range. 
The agreement is between 
$0.5 -2 \mu$ Hz. At high frequencies, the agreement is less, due to the description 
of the outer layers. See Turck-Chi\`eze et al. (1997) for a discussion 
on absolute frequency values for different solar models.
If one compares our calculations with and without diffusion, we note 
an improvement of 4-5$\mu$Hz at low order for models incorporating microscopic diffusion,
\ due to a higher helium and Z element composition in the region of the sun where radiation dominates, and also a slight improvement for the fine spacing value $\bar{\delta}_{0,2}$.\\ 
On the other hand, the higher abundances of He and metals in the solar core, increase 
the discrepancy with neutrino experiments ($^{71}$Ga +7\%, $^{37}$Cl +27\%, $^8$B +32\%). 

\noindent
e) a different shape of figure 4 in comparison with previous studies (Basu et al. 1997, 
Turck-Chi\`eze et al. 1997), due to the improvement in modeling the physical processes,
specifically the introduction of the OPAL96 opacity coefficients,
including 19 elements and the updated reaction rates. 
This shows the importance of a proper treatment of  
the known physical processes at the level of accuracy we  can reach today to be able to interpret the
residual difference.
  
  \noindent
  f) For the neutrino predictions, we use the recent work of Bahcall et al. (1996) for the updated 
  absorption cross sections. Compared to the previous predictions, our diffusive model (see table 4) 
  does not 
reach such large neutrino flux predictions due to the reestimate of the nuclear reaction 
rates which leads to
a reduction of the neutrino fluxes and practically a compensation of the microscopic 
diffusion effect.\\ 

\section{NEW CONSTRAINTS ON SOLAR MODELS FROM RECENT SEISMIC RESULTS}

The recent sound speed profile, deduced from the helioseismic experiments
 leads to a  characteristic deviation 
with standard solar models. It shows a bump near 
the base of the
 convection zone and a slower sound speed in the nuclear core with the transition  close 
 to the $^3$He production zone 
 (i.e $\sim$ 0.28 R$_{\odot}$) (see Turck-Chi\`eze and Brun 1997) or near the edge of the nuclear core.
  The physical processes 
 which are at the origin of this behaviour, must be investigated in order to improve 
  the description of the solar interior and also on the prediction of the observables.
 In a previous paper (Turck-Chi\`eze et al. 1997), we have shown the sensitivity of 
 this shape to some ingredients of the calculation.
  
 In this section, we check first the consequence  of the
 known uncertainties of the physical processes already included in our 
 reference solar model on the sound speed (see section 2). Then we  discuss 
  the possibility of some mixing in the core (sometimes evoked to decrease 
  the central temperature), and 
  some specific points as  the actual knowledge of the solar radius. 
  
The main bump, close to the convective-radiative transition zone is of great interest, 
as it could  certainly be the first manifestation of hydrodynamical phenomena. 
Coupled with the rotation 
profile in this range it is a strong constraint on tachocline instabilities inside the Sun 
(Zahn 1992, Spiegel and Zahn 1992, Kosovichev et al. 1997, Corbard et al. 1998). Nevertheless, the detailed interpretation of this peak needs 
to review
also the influence of the quality of some opacity coefficients and 
the knowledge of the abundances. This concerns mainly the oxygen and iron elements.
A preliminary study has been done by Turck-Chi\`eze et al. (1997), which show that  
the bump may 
be reduced by a localized $5\%$ variation in the opacity coefficients, which is certainly 
the order of uncertainty we may have on this quantity at the  radiation-convection edge. Moreover, such a study 
shows that 
the effect is local and has no consequence on the central part of the Sun. This region 
will be extensively studied in our next paper " II The radiation convection transition".

 As the knowledge of the nuclear part of the Sun 
is the main objective of this study,  we shall  concentrate 
on the  radiative zone in this paper.

\subsection{The role of the uncertainties on opacity coefficients and heavy element
compositions}

Considering the broad deviation in the intermediate region between the nuclear region
and the convection zone, and noting the improvement done by the microscopic diffusion, 
which modifies the composition across the sun, we first investigate a possible 
solution of the discrepancy based on an
 opacity increase compatible with the abundance uncertainties on
  the C, N, O, Fe elements.
  In fact, their ionization states play a crucial
    role in the opacity coefficient in the radiative zone. Therefore, in order to 
    simulate
    an opacity correction connected to the treatment of the heavy elements 
    in the absorption 
    opacity coefficients or to the knowledge of their abundances, 
    we fit a correction to $\kappa (T, \rho, X, Y,Z)$ compatible with Figure 1 
    of Courtaud et al. (1990), which  represented the proportion of 
    the heavy element opacity to the total opacity. This curve has 
    been established for 
    an iron composition about $30\%$ greater than we believe today.
   In fact, 
in current tables the C, N, O, Fe elements contribute about 70$\%$ of the opacity near the base of the convective zone (Richer 1997, Richer et al. 1998). 
 The correction we applied is of 1.5 
    $\%$ in the center and $5\%$ at the base of the convective zone. The results appear in table 6 for the 
    structural quantities and figure 5a) for the sound speed. 
    We note that the sound speed matches the helioseismic data better, 
    ($\delta c^2/c^2 < 0.6\%$, $^4$He$_{surf}$=0.248 and R$_{bcz}=0.712$), 
    without changing greatly the neutrino flux predictions.
     Also, as expected, the predicted acoustic mode frequencies  
     match closer the GOLF frequencies (see figure 8).

This model (Opac Inc) shows the sensitivity of the Sun structure to localised opacity 
changes (see also Gabriel 1997) and the necessity to improve our knowledge on the 
radiation-matter interaction as well as on chemical element 
abundances.

\begin{figure}[htbp]
\vspace{2.cm}
\setlength{\unitlength}{1.0cm}
\begin{picture}(7,15)
\includegraphics{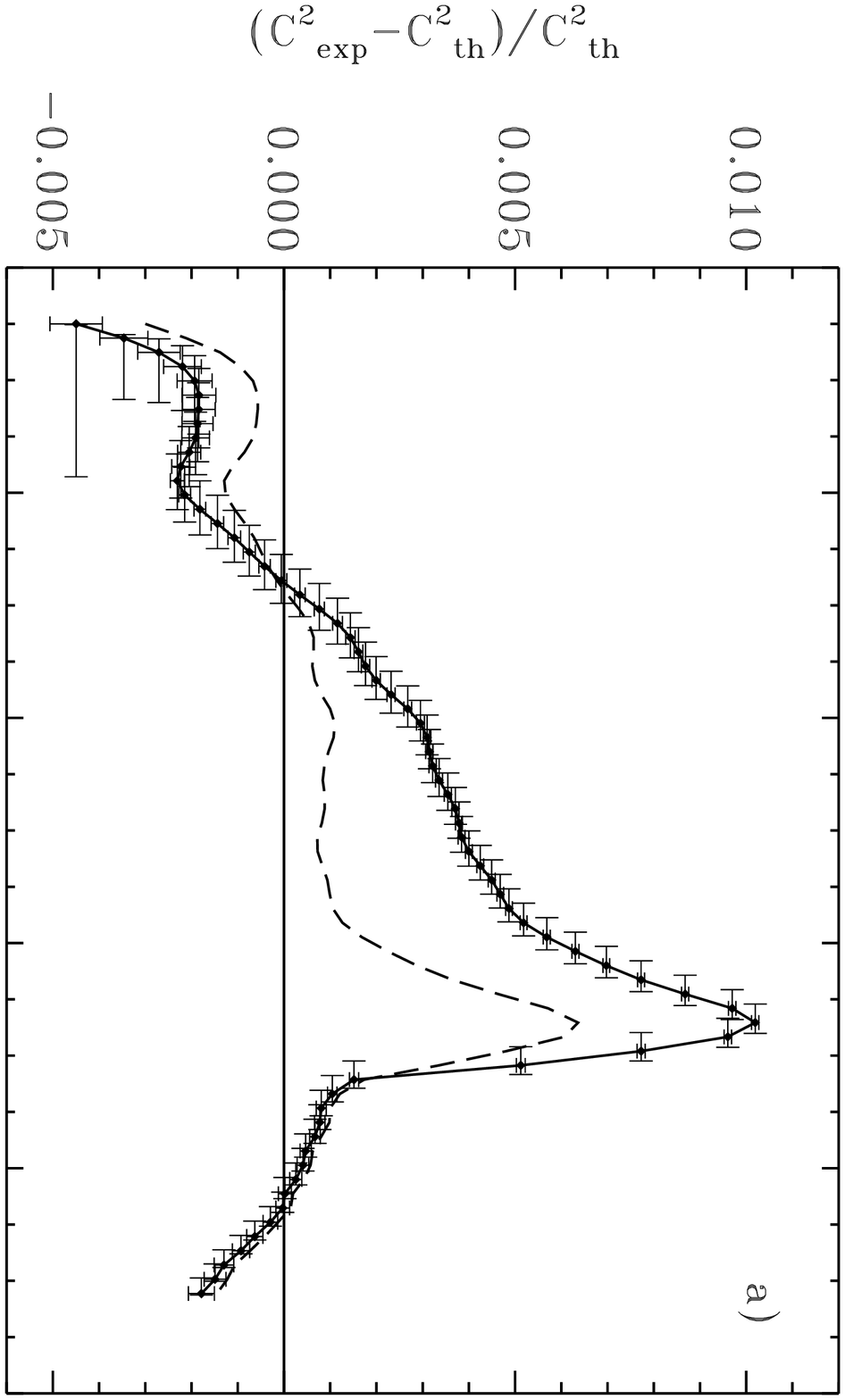}
\includegraphics{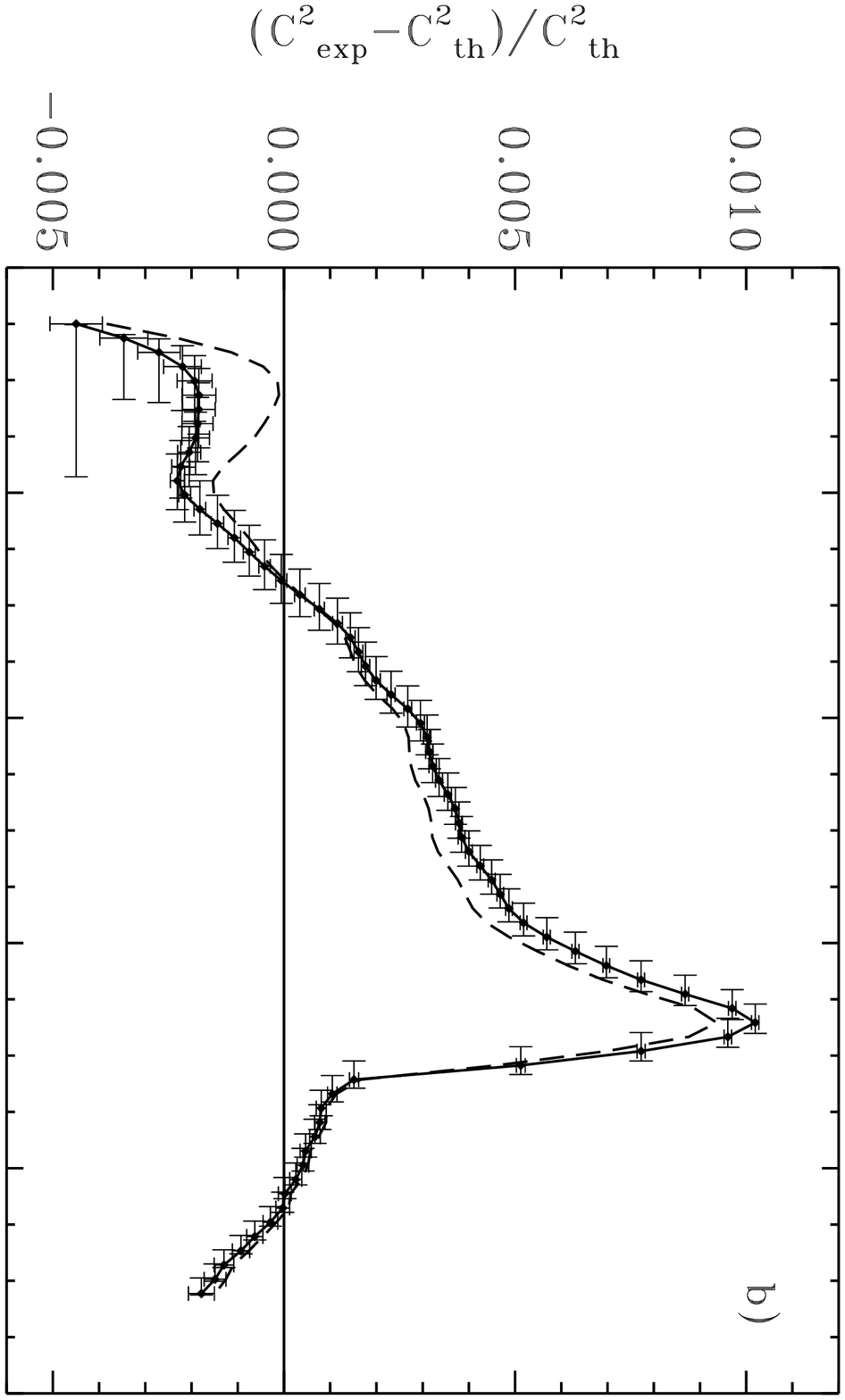}
\includegraphics{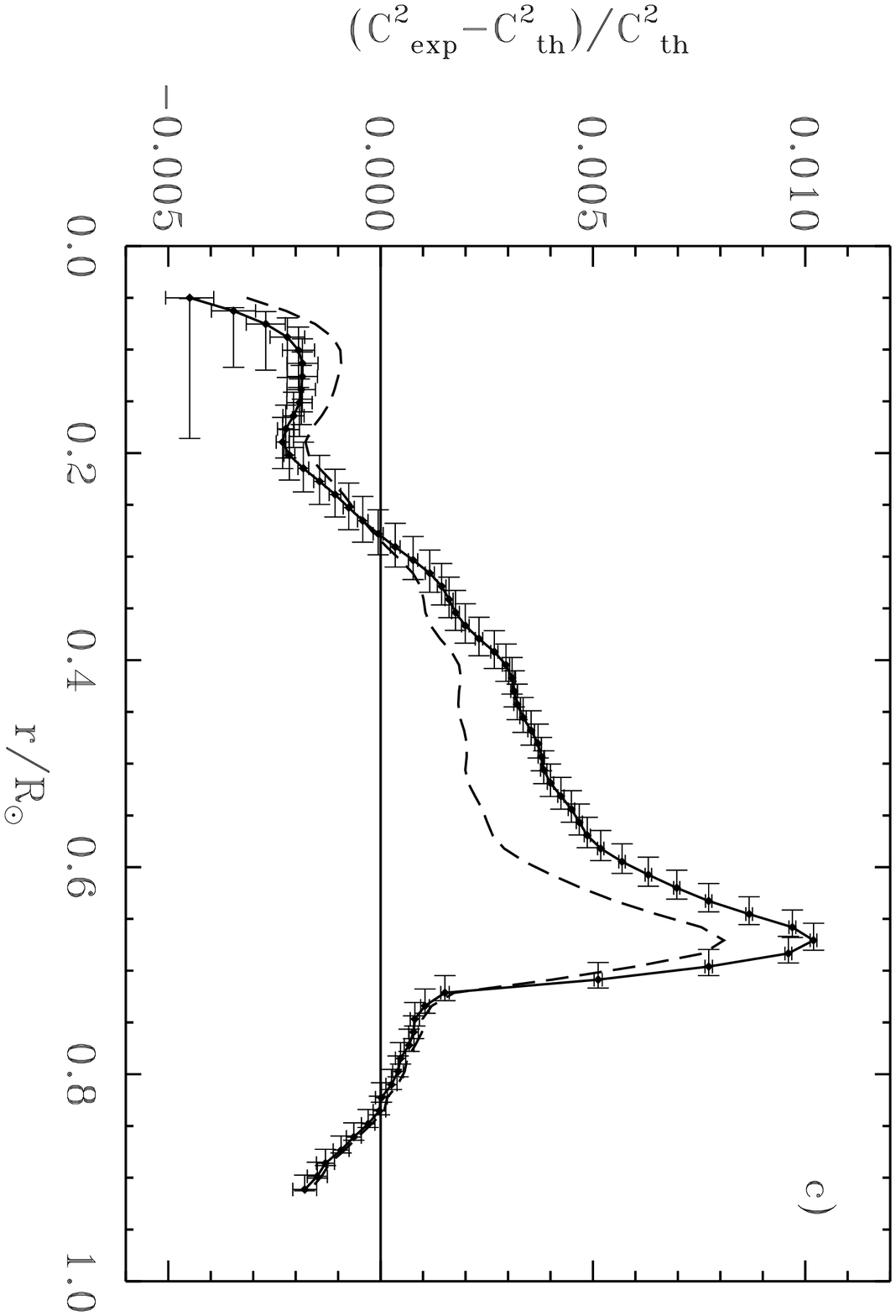}
\end{picture}
\vspace{-0.2cm}
\small
\begin{center}
Fig 5: Sound speed square difference between the Sun seen by GOLF +LOWL
 (see Turck-Chi\`eze et al. 97) and models. Full line  standard model with 
 experimental seismic error bars. Dashed line:  a) a model with an increase of opacity between 1\% in the central region up to
  5\% at the bottom of the convective zone, b)  a model taking into account 
 the uncertainty on the nuclear reaction rates to define a ``minimal nuclear" model, c) a model which takes into account 
 the uncertainty on the microscopic diffusion, here the velocities are increased by 15\%
 in comparison with our standard model.
 \end{center}
  \normalsize			
\vspace{-0.3cm}
\end{figure} 

\begin{table*}[htbp]
\begin{center}
  \caption[]{\label{table 6}     
Solar models including a modification of 
the basic physical ingredients inside the present uncertainties. 
Same variables than in table 5.}     			
\end{center}
\begin{center}
   \begin{tabular}{p{3cm}*{4}{c}}  	
    \hline
     Parameters	 & \multicolumn{3}{c}{models}\\
      & Opac Inc & Min Nuc  &   Microsc. Inc     &\\
    \hline
     Opacities & OPAL/K 5800& OPAL/K 5800& OPAL/K 5800&\\
     Diffusion  & yes   &    yes  &  yes \\ 
    
     $\alpha$  & 1.84  &   1.85 & 1.86\\
     Y$_0$  &  0.277  &   0.273  & 0.274 \\
     Z$_0$  &  1.95 $10^{-2}$& 1.96 $10^{-2}$ & 1.99 $10^{-2}$\\
     (Z/X)$_0$ & 0.0277  &  0.0277 & 0.0283\\
     \\
     Y$_s$     &   0.248   &  0.244    & 0.240\\
     Z$_s$  &   0.01799   &  0.01809   &    0.1817 \\
     (Z/X)$_s$ &  0.0245     & 0.0245     &  0.0245\\
     $\tau_b$  &     2   &   2    &  2 \\
     \\
     R$_{bcz}$/R$_{\odot}$    &  0.712    &    0.715   & 0.713\\
     T$_{bcz} \times 10^6$    &    2.208   &  2.178   & 2.188 \\
     Y$_c$ &  0.640   &   0.631  &  0.638\\
     Z$_c$ &   0.0207   &  0.0208   &  0.0213\\
     T$_c \times 10^6 (K)$   &  15.72    & 15.61  & 15.70\\
     $\rho_c$ (g/cm$^3)$   &    152.63    &  150.64    &  152.44\\
     \\
     $\rm ^{37}Cl $ (SNU) &  7.55  & 5.19    &  7.42\\
     $\rm ^{71}Ga $ (SNU)  &  129.0    &  119.0   &  128.4\\
     $^8$B    (10$^6$/cm$^2$/s) & 5.10      &   3.21  &  5.00\\
    \hline
   \end{tabular}
  \vspace{-0.3cm}
  \end{center}
\end{table*}

\subsection{The role of the uncertainties on the nuclear reaction rates}

The great sensitivity of the  neutrino fluxes produced 
by the interaction of electron and proton with the $^7$Be
to the different nuclear reaction rates has been largely discussed in the past  (Turck-Chi\`eze and Lopes 1993, 
Castellani, Degl'Innocenti and Fiorentini 1993) and has stimulated 
new measurements and theoretical efforts (see section 2). On the other hand 
the sensitivity 
of the sound speed to these important ingredients 
of the solar modeling 
is not so large, except for the pp
reaction rate, as it modifies  the energy production directly
(Dzitko et al. 1995, Turck-Chi\`eze et al. 1997).

Here we discuss a ``minimal nuclear solar model'' (Min Nuc), 
where we modify the nuclear cross sections  
within the experimental error bars of the new compilation
in order to get
the minimal neutrino fluxes (Dzitko et al., 1995). The modifications are:
 +2.2\% on $^1$H(p,e$^+$ $\nu$)$^2$H, +8\% on
$^3$He($^3$He,2p)$^4$He, -10\% on
$^3$He($\alpha$,$\gamma$)$^7$Be 
and -20\% on $^7$Be(p,$\gamma$)$^8$B. The results are summarized in table 6 
and figure 5b.
They show a slightly better agreement with the present helioseismic 
results in the solar core and a reduction 
of the neutrino fluxes  by -6.5 \% for $^{71}$Ga,  -28\%  for $^{37}$Cl,  -33\% for $^8$B at $1 \sigma$ level, 
which are, except for gallium case, half the discrepancies between neutrino detections and predictions.
The corresponding  acoustic modes frequencies are modified by less than 0.2$\mu$Hz 
(see figure 8). 

One may add some comments on the CNO uncertainties which are rather large: (p, $^{12}$C): 15\%,
(p, $^{13}$C): 13\%, (p, $^{14}$N): 11\%, (p, $^{16}$O): 18\%. As the CNO contribution to the total
luminosity is rather small, 1.26\% in our standard model, an error in these cross sections has a 
small impact on the neutrino predictions:
about 11\% luminosity modification of this small part has an effect of 
0.7 SNU on gallium 
prediction, 0.11 SNU on chlorine and 
practically no effect on the boron flux. On the other hand, if we try to minimize the neutrino fluxes
on gallium and chlorine detectors by reducing the CNO contributors, this can lead up to an added -2 SNU on gallium 
and -0.5 SNU on chlorine at the 1 $\sigma$ level.

We notice that the sound speed in the nuclear range may be a good constraint 
on the pp reaction rate
which is not  measurable in the laboratory due to its very small cross section.
But one must not  forget that helioseismology has at present difficulty in discriminating between different 
sources of uncertainties. 
In the next section, we discuss the effect of these modifications on the g-mode predictions.

One finally observes that the $^8B$ neutrino flux and the $^7Be$ neutrino flux, 
respectively peaked at 0.05 and 0.1 R$_{\odot}$ are not well constrained 
due to the present  seismic 
uncertainties (see figure 5). A slightly smaller temperature in the very 
central region is not excluded.

\subsection{The role of the uncertainties on the effect of the microscopic diffusion}

As it has been mentioned previously, one may question the way we introduce the microscopic diffusion 
velocities 
for two reasons: first because we use approximate formulae at about 5-10 $\%$ accuracy, secondly 
because the settling of the chemical species is inhibited by turbulent diffusion 
(Proffitt and Michaud 1991, Richard et al. 1996, Gabriel 1997). Therefore,
in this section we look to the sensitivity of the results to a variation of the microscopic diffusion
velocity.  We have first considered a model where we have increased the helium diffusion velocity by 8\% 
and the heavy elements diffusion velocity by 15\% in order to simulate a more 
complete microscopic calculation.
A improvement of the sound speed by no more than 0.1 \% occurs with a slight increase of the neutrino fluxes
and a decrease of the surface helium composition to 0.241, which seems to be a 
rather small value. Then to increase 
the effect on the sound speed,
we have considered a model where both the helium and the heavy element microscopic diffusion are 
increased by 15\% (Microsc Inc). In these conditions, the effect on the sound speed is 
noticeable 
(see figure 5c), 
but, as seen in table 6,
the reduction of the photospheric helium is important and difficult to reconcile
 with the helioseismic photospheric value  except if some turbulent mixing at 
 the base of the convective zone is evoked to reduce the element settling. 

Finally, we notice a slight improvement in the sound speed when we increase the microscopic diffusion, but it seems that
such increase is not really supported by the global observations.

\subsection{Other questions?}

\subsubsection{Are the helioseismic results compatible with $^3$He turbulent diffusion ?}

In order to try to solve the neutrino problem and to understand the weakest points
 of classical stellar evolution in solar-like stars as the Li and Be abundances or the evolution of the 
 rotation along the stellar life, much theoretical works have been devoted 
 to the study of possible instabilities inside the stars (Schatzman and Maeder 1981, 
 Lebreton and Maeder 1987, Zahn 1992, Morel and Schatzman 1996,
 for a review and further references: Turck-Chi\`eze et al. 1993).
  Some of them have been rejected by physical 
 considerations, such as the partial inhibition of turbulent mixing by the gradient of 
 composition. Some others have proposed very recently such as mixing due to gravity waves (Schatzman 1993,
  Kumar et al. 1997, Zahn, Talon and Matias 1997).  An additional
 question is the reaction of the $^3$He peak located at the boundary of the nuclear core at 0.3 $R_{\odot}$
 (see figure 5 of Turck-Chi\`eze and Brun 1997).
 
 The solar angular velocity profile would help in this discussion but this 
 quantity is difficult to extract below $0.4 R_{\odot}$
 due to the small value of the splitting  of the acoustic modes. It is typically 
 4 times the estimated error bar and 
 largely influenced by the inherent effect of the stochastic excitation. Some tentative
  estimates which appear 
 partly contradictory, exist (Elsworth et al. 1995, Lazrek et al. 1996) . 
 With very continuous data
 produced by SOHO helioseismic experiments (GOLF, VIRGO and MDI), 
we begin  to make progress on the accuracy of 
this rotational splitting (Lazrek et al. 1997).
We hope to have a second constraint from the results of the  COROT mission 
(Catala et al. 1995), an asteroseismic 
project which will be launched in 2002, through measurements of the rotational splitting
of young clusters.

In the absence of definitive constraint on the internal rotation of the Sun, 
we  limit  our investigation to 
the profile of the 
well known sound speed, in calculating ``ad hoc'' models which may simulate 
any reasonable mixing.
Following the work of Lebreton and Maeder (1987), we have computed several 
models which include a turbulent 
term  $D_T=\nu_{rad}*Re^*$. We have used different critical Reynold numbers, 
$Re^*$ =\,cte = \,20 or $Re^*= \,20 \exp^{-1/2(r-a/\sigma)^2}$ with $a = 0.3$ 
or $0.71$.
We have also tried an other expression for the turbulent term, proposed by 
Morel and Schatzman (1996) to describe the effect of internal waves:

  $$D_T=D_0*\exp^{-1/2(r-0.22/\sigma)^2}  (2)$$
  
\noindent  
   with $D_0$ values varying from 10$^2$ to 
 10$^4$ cm$^{2}$ s$^{-1}$ and $\sigma$ one from 0.02 to 0.04.
In the different cases, the photospheric helium is slightly increased, the 
neutrino flux on chlorine detector or water 
never reduced by more than 15$\%$, the sound speed profile never improved and is 
even worst (e.g. $\delta c^2/c^2 > 2\%$) (see also Richard and Vauclair 1997). 
It seems to show that the present results do not support  evidence for mixing today 
(see figure 6 for a typical example). 

\begin{figure}[htbp]
\setlength{\unitlength}{1.0cm}
\begin{picture}(10.5,7.5)
\includegraphics{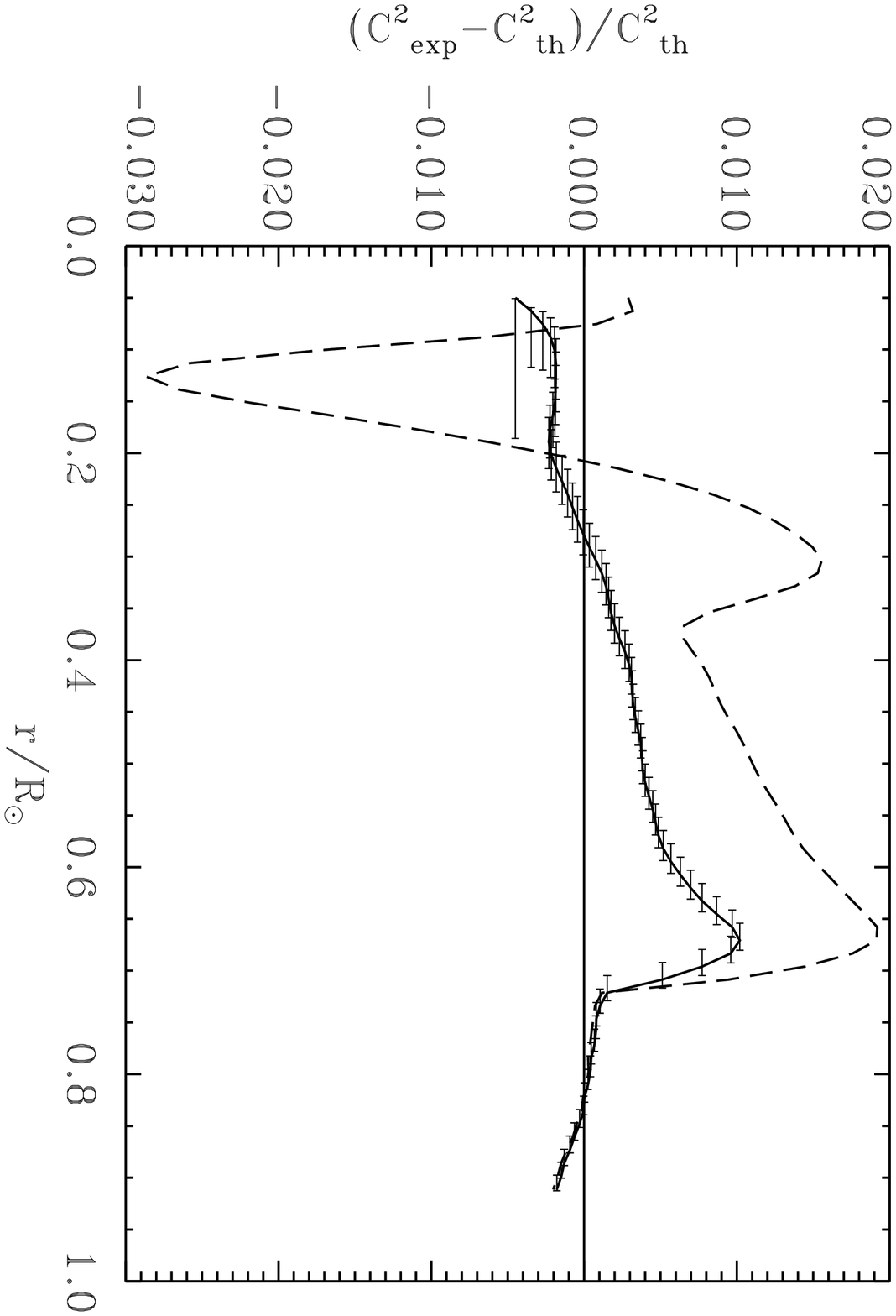}
\end{picture}
\vspace{-0.2cm}
\begin{center}
Fig 6:	
 Sound speed square difference between the Sun seen by GOLF +LOWL
 (see Turck-Chi\`eze et al. 97) and a model where we have introduced 
 a small amount of mixing of the elements 
 following expression (2) with $D_0$ = 1000 cm$^{2}$ s$^{-1}$.
\end{center}			
\vspace{-0.3cm}
\end{figure}   

Of course this is not a definitive conclusion as we cannot exclude 
a temporal dependence of this kind of coefficient.
A more general study must be supported by theoretical 
investigation of the evolution of the angular momentum
with time.

\subsubsection{ What is the effect of the solar age?}

The present accuracy of the determination of the sound speed is so impressive that we have noticed
different results for very similar physics included in standard solar models.
This must be solved in the near future if one hopes to progress on the interpretation of the 
seismic tool. Here we question the problem of the age of the Sun where we adjust
the present luminosity and radius. Figure 7 illustrates this point in showing two models 
converged at two different 
ages: 4.52 and 4.6 Gyr (also currently used), starting at the begining of hydrogen burning. 
This increase leads to an increase of the neutrino predictions by less than 2.5\% but  the 
effect is  not negligible on the sound speed. 
This point must be taken into account in a comparison 
between several models which include premainsequence or not and use slightly different solar ages.
Is it a clear signature that the Sun could be a little older than one generally thinks?

\begin{figure}[htbp]
\setlength{\unitlength}{1.0cm}
\begin{picture}(10.5,7.5)
\includegraphics{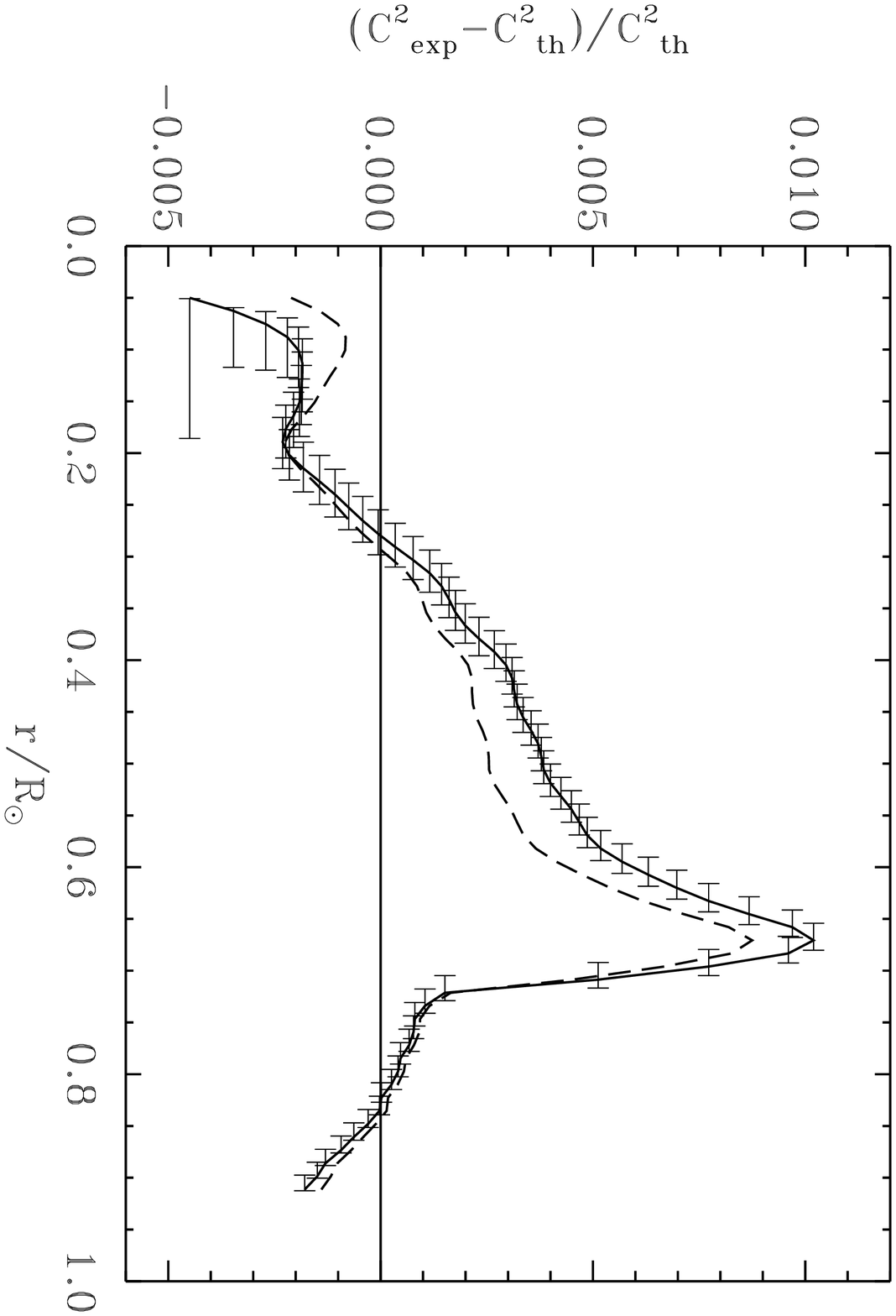}
\end{picture}
\vspace{-0.2cm}
\begin{center}
 Fig 7: Sound speed square difference between the Sun seen by GOLF +LOWL
 (see Turck-Chi\`eze et al. 97) and two models converged at different solar age: 
 our standard model at 4.52 Gyr (full line)  and  a model converged at 4.6 Gyr 
 (dashed line). 
\end{center}			
\vspace{-0.3cm}
\end{figure}  

\subsubsection{ Is the solar radius well determined?}

Due to the quality of the helioseismic measurements, one generally adjusts 
the solar radius at
the estimated value with an accuracy of some $10^{-5}$. This is useful
 for the determination of 
the absolute value of the acoustic wave frequencies which vary as $R_{\odot}^{-3/2}$. 
If we believe the present inversions, we may question the slope
observed just above the base of the convective zone (see figures 5a, b and c), 
where  the temperature gradient 
is purely adiabatic. So to verify this point, we have modified the radius 
by -300 km to see how the agreement will
be modified following the recent possible reestimates of the photospheric solar radius 
(Schou et al. 1997, Antia 1998, Brown and Christensen-Dalsgaard
1998). In this case we have found a more constant agreement 
just above the edge of the convective zone
even if 
the absolute difference remains 0.1\%. The effect on the frequencies 
is a reduction of 2 $\mu Hz$
 at 3 mHz and -0.64 $\mu Hz$ at 1 mHz.
This point could contribute to the general trend observed between measurements 
and absolute predictions 
(Turck-Chi\`eze et al. 1997).

 \vspace{0.6cm}

\section{DISCUSSION}

In the previous section, we have discussed several modifications
of our reference model to see which  may better match 
the present acoustic mode observations, in varying the main physical 
processes inside their uncertainties, but also evoking  
some other processes, 
such as mixing in the core. We consider 
that  the two modified 
models: changing the opacities in the radiative region or the nuclear reaction rates,
inside the known error bars, match
better  the present helioseismic data, one in the intermediate region between 
the nuclear one and the convection zone, the other in the solar core. Consequently,
 we  discuss 
these two models in details.

\subsection{Influence of the physical processes on the acoustic mode predictions}
In figure 8, we present the  acoustic mode frequency differences between these two models 
and our reference model for l= 0 and 2. 
We note that the GOLF results of the first year agree rather well with the model 
\begin{figure}[htbp]
\setlength{\unitlength}{1.0cm}
\begin{picture}(10.5,7.5)
\includegraphics{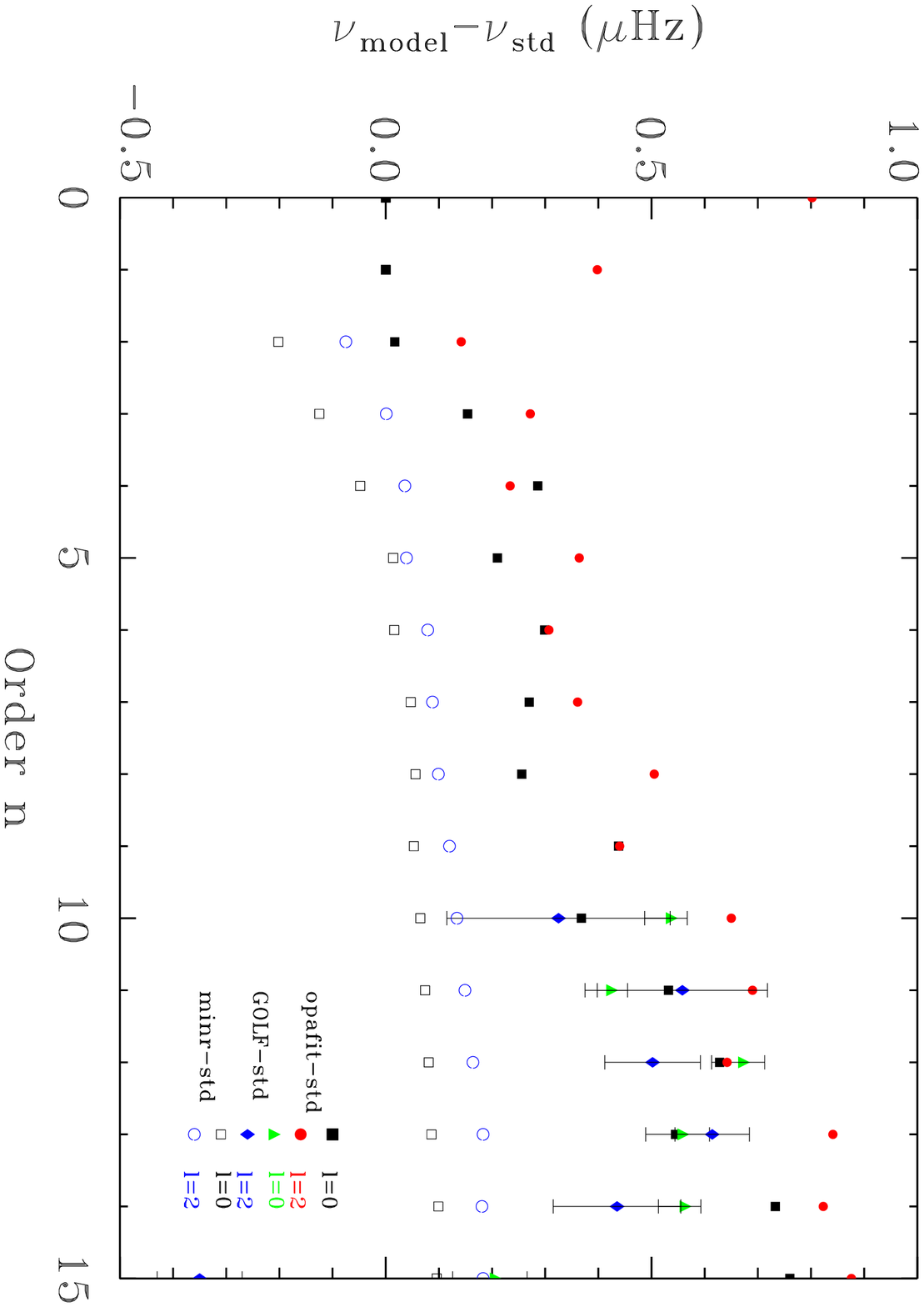}
\end{picture}
\vspace{-0.3cm}
\begin{center}
Fig 8:	
Difference between theoretical acoustic frequencies of various models 
(the reference being the standard diffusive model), {\it full} symbol for the model 
with a modification of opacity, {\it empty} symbol for the model with 
a modification of nuclear reaction rates. 
Superimposed  GOLF data set taken between 11 April to 25 August 1996 
(Lazrek et al. 1997).
\end{center}			
\vspace{+0.5cm}
\end{figure}

where the opacity coefficients have been slightly modified. 
In the case of the modification of the nuclear reaction rates 
the best agreement in the solar core, observed on the sound speed, 
is not visible in  the absolute values of the 
acoustic modes as they are influenced very little by the solar core.


\subsection{The corresponding gravity mode predictions}

It is clearly established that the gravity modes are a better probe of the nuclear core. 
Table \ref{table 7} 
shows some theoretical predictions of these modes, obtained with the pulsation code of 
Christensen-Dalsgaard (1982) for our reference model. In this table, we concentrate
 on the most observable 
gravity modes (e.g l$=1$ and l$=2$) which may be accessible to the satellite SOHO. 
 Following the first asymptotic 
approximation, the gravity frequency is proportional to $I= \int_0^r{N/r dr}$,
 where N is the Brunt V\"ais\"al\"a frequency ($ \rm N^2 = g \large( {1 \over \Gamma_1}
 {d ln P \over dr} - {d ln \rho \over dr} \large)$).
 60\% of 
 the frequency of the gravity modes is built on the value of the Brunt V\"ais\"al\"a frequency in
  the inner 0.2 $R_{\odot}$.

\begin{figure}[htbp]
\setlength{\unitlength}{1.0cm}
\begin{picture}(10.5,7.5)
\includegraphics{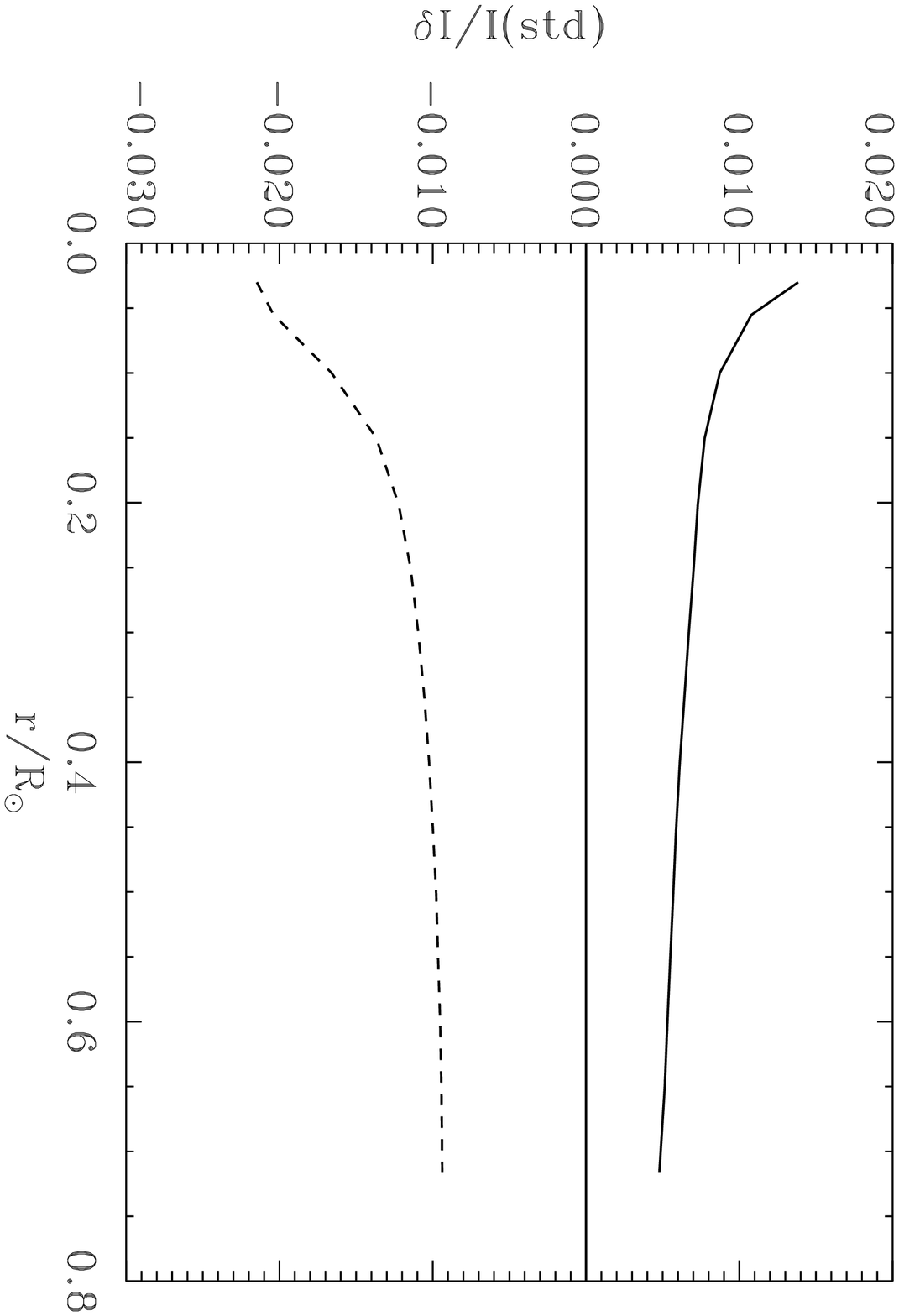}
\end{picture}
\vspace{-0.2cm}
\begin{center}
Fig 9:	
Difference of the radial dependence of the g mode frequency 
for the model with an opacity change (full line) and the model called "Min Nuc"
(---) in comparison with the standard model, in the first asymptotic approximation. 
\end{center}			
\vspace{-0.3cm}
\end{figure}   

 Therefore in figure 9, 
  we show the comparison of the convergence 
  of this integral
  for our two modified models in comparison with our reference model, to better 
  understand how a physical modification of the solar model influences the gravity modes.

\begin{table*}[htbp]
\begin{center}
\caption[]{\label{table 7} Gravity mode frequencies ($\mu Hz$)
obtained from our standard model, with the pulsation code of Christensen-Dalsgaard (1982).}
\vspace{0.5cm}
\begin{tabular}{p{1cm}*{3}{c}}
\hline
   n  &    $l=1$ &     $l=2$  \\
   1  &   260.3  &  294.5 \\
   2  &  189.3  & 254.0  \\
   3  &  151.6 & 220.2 \\
   4  &  126.3 & 192.2  \\
   5  &  107.9  &   168.6  \\
   6  &  94.31  &  149.5  \\
   7  &  83.68  &  133.9  \\
   8  &  75.21  &  121.1  \\
   9  &  68.23  &  110.4  \\
  10  &  62.45 & 101.4  \\
 \hline
\end{tabular}
\end{center}
\end{table*} 

\begin{figure}[htbp]
\setlength{\unitlength}{1.0cm}
\begin{picture}(10.5,7.5)
\includegraphics{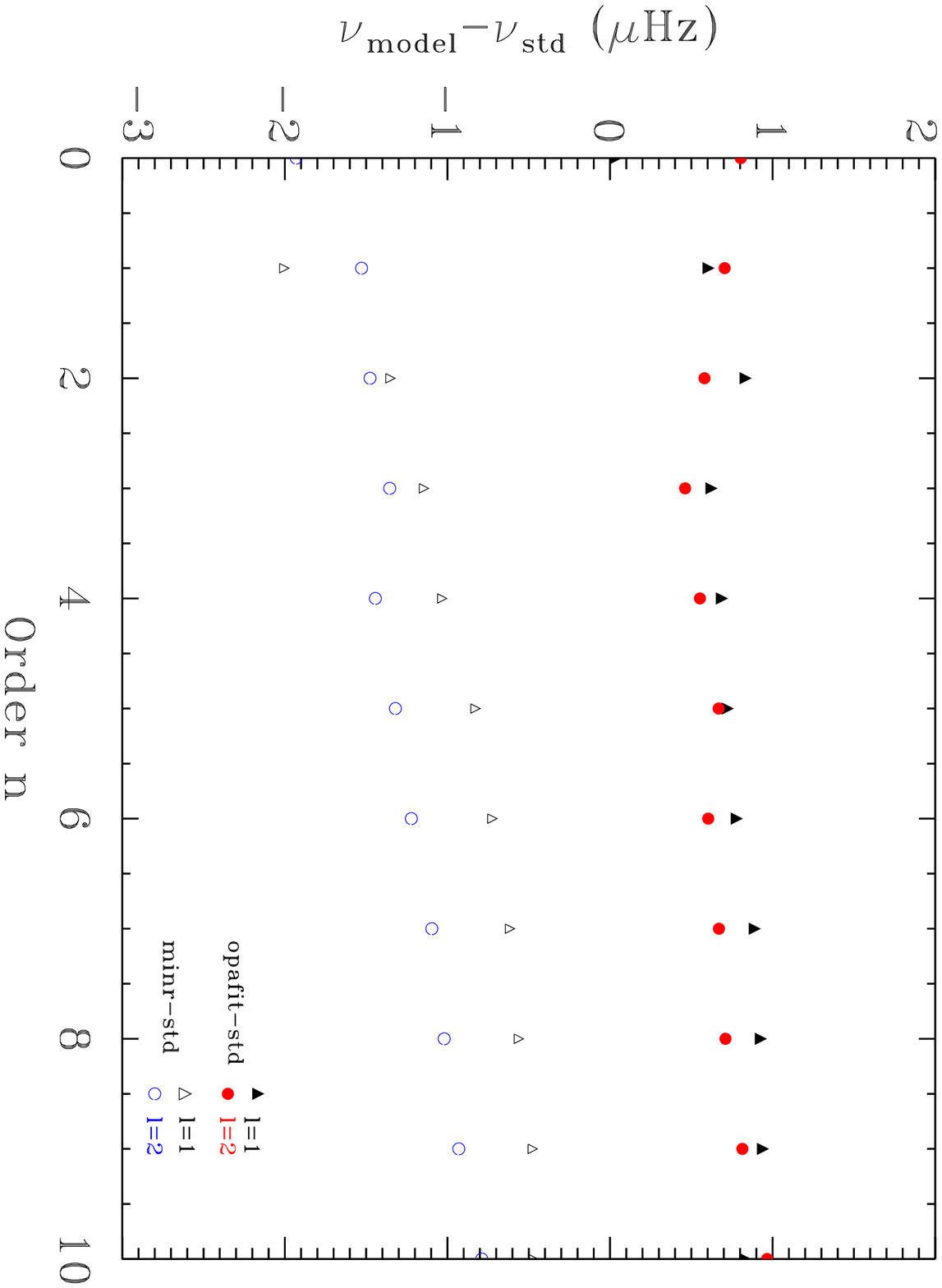}
\end{picture}
\vspace{-0.3cm}
\begin{center}
Fig 10:	
Difference between the gravity modes of the "Opac Inc" model {\it full} symbol  or of the "Min Nuc" model {\it empty} symbol and 
those of our standard solar model.
\end{center}	 			
\vspace{-0.5cm}
\end{figure}

We then calculate the g-mode frequencies for these different models with 
the pulsation code.
The modification of the nuclear reaction rates introduced in the model "Min Nuc"
does induce a decrease of frequency  which may reach 2$\mu Hz$ 
in the range between 50-300 $\mu Hz$ (figure 10). This illustrates that the gravity modes
 are four to ten times more sensitive to the solar core modifications 
 (see also Brun and Turck-Chi\`eze 1997) than acoustic modes. 
 We also notice that a slight opacity increase  and  
 the modified nuclear reaction rates, change the gravity mode frequencies in the 
 opposite direction. 
 These variations and their dependence in frequency, are substantially above the 
 experimental resolution and could be measured if the gravity modes maybe identified at low 
 order. One may also remark that the  frequencies dependence of the gravity modes on physical processes is not easy to anticipate
 and contribute to the difficulty of the identification of such modes.
As the opacity change acts in the opposite way compared to the nuclear modified model 
for the central thermodynamic quantities, but not for $\delta c^2/c^2$, or $^4$He$_{surf}$ 
and R$_{bcz}$,
this emphasizes the fact that a coupled analysis of acoustic modes 
and gravity modes  could be useful to 
differentiate different physical processes.

\section{CONCLUSION}
 In this study of the radiative region of the Sun, we have observed that the standard picture of the stellar evolution 
 is very robust when challenged by the present seismic constraints, if one accepts 
 the modification of some physical processes such as opacity or nuclear reaction rates
 inside their inherent error bars. This is a very important issue. We have proposed two models 
 which may better match the information extracted from acoustic mode measurements, in order to improve
 the standard observable predictions: acoustic modes, gravity modes and neutrino fluxes.
 
 The present seismic data do not support clearly any mixing in the core: 
 effectively, any tentative model of mixing does not improve the sound speed comparison.
 This fact does not allow us to say that there is no mixing in the solar core 
 as one can imagine compensation effects
 or mixing in the past, all of them cannot be detected on the present profile. 
 We consider 
 only that the present 
 helioseismic sound speed profile does not allow us to deduce any mixing 
 characteristics.
 A detailed analysis of the radial rotation profile coupled 
 to a theoretical time dependent
 sound speed profile is necessary before conclusion on this specific point.
 
 Even if the present solar structure seems rather well under control,
  we have noticed  that our solar neutrino predictions are significantly
   smaller than  recently published ones. 
  After a comparison of the used ingredients, we can say that 
  this discrepancy is mainly due to the reestimate of 
  the nuclear reaction rates. Moreover, a persistent large uncertainty 
  in chlorine and water detector 
  neutrino predictions exists. A 30 \% reduction of the predicted values 
  is not ruled out and even favoured by the present study. 
  
  It seems that the present seismic results 
   contrain more and more the pp reaction rate and could favour a slightly higher value
   compatible with the estimated error bar. It is possible
    to believe that one begins to be sensitive to the 
   effect of the $(^3He, ^3He)$ interaction (shape of the sound speed at low radius). But we also note that the solar structure 
  in the inner 0.1 R$_{\odot}$ is not yet constrained by the present seismic results.
  This confirms 
 that
  the sensitivity of the present helioseismology (sound speed and acoustic mode analysis)
  to 
  the nuclear reaction rates stays rather low. 
 Moreover the sensitivity to 
 all the assumptions on screening or reaction rates of the pp chain II or III
 or CNO cycle is still rather poor (see Turck-Chi\`eze et al. 1997). 
 We can then conclude that the properties 
 of the internal plasma is not checked by present seismology results 
  (r $<0.2$ R$_{\odot}$). This point can only be checked by improved nuclear 
 experiments at low energy or maybe by high energy laser experiments which are
  presently under investigation in our laboratory.
 
 In parallel, the knowledge of the internal rotation and of the g-mode
  frequencies will contribute toward the knowledge of the nuclear core and
   to solve some issues that the 
  classical stellar evolution has not totally answered such as 
 the  photospheric lithium or the history of the angular momentum of the Sun.

We have tried to determine some standard models (inside the known error bars of the ingredients),
 in better agreement with 
the seismic constraints. In doing such models, we constrain, better than previously, 
the initial helium content. 
This study supports a value between 0.273-0.277 in mass fraction.

As our accuracy progresses, we must also  be more careful with
 the available information. We have observed  different shapes 
of the sound speed in the core over the last 2 years, supported
 by different observations or 
different physical processes introduced in the calculation. We notice 
that the solar age plays a non negligible effect.
The next  years are absolutely crucial with all experiences of 
solar neutrino detection or seismology
in operation. Moreover, due to the extreme accuracy we reach now, 
the comparison between  models of different groups will contribute 
to definitively fix what we know on the present solar core.

\vspace{0.5cm}

\acknowledgments{We would like to thank J. Bahcall, J. Christensen-Dalsgaard, G. Michaud, J. P. Zahn
and our colleagues on the GOLF team for very useful discussions 
and, for interactions in our common work. We are particularly grateful to S. 
Tomczyk for giving us access to the LOWL acoustic mode frequencies and to S.
Basu and J.
Christensen-Dalsgaard for their work in the inversion of 
the GOLF and LOWL data and for the use of the pulsation code.}


\end{document}